%% file: main.tex
\documentclass[a4paper,USenglish]{lipics}

\usepackage{microtype}

\usepackage[inline]{enumitem}

\usepackage{tikz}
\usetikzlibrary{calc}
\usetikzlibrary{positioning}
\usetikzlibrary{shapes}
\usetikzlibrary{fit}
\tikzset{anonymous/.style={draw, circle, inner sep=0.7mm}}
\tikzset{forbidden/.style={draw, rectangle, inner sep=1mm}}
\tikzset{necessary/.style={draw, isosceles triangle, isosceles triangle apex angle=60, shape border rotate=90, minimum width=2.5mm, inner sep=0.5mm}}

\usepackage{circle} 
\newcommand{\medcirc}{{\Circle}}

\newcommand{\decisionProblem}[3]{%
\begin{center}
\fbox{%
\begin{minipage}{.95\linewidth}%
#1
\begin{itemize}[topsep=1mm,itemsep=1mm]
\addtolength{\leftskip}{11mm}
\item[Input:] #2
\item[Question:] #3
\end{itemize}%
\end{minipage}%
}%
\end{center}
}

\renewcommand{\O}{\ensuremath{\mathcal{O}}}

\renewcommand{\phi}{\ensuremath{\varphi}}
\newcommand{\prob}[1]{\textsc{#1}}
\newcommand{\probSS}{\prob{Secure Set}}
\newcommand{\SSF}{\prob{Secure Set\textsuperscript{F}}}
\newcommand{\SSFN}{\prob{Secure Set\textsuperscript{FN}}}
\newcommand{\SSFNC}{\prob{Secure Set\textsuperscript{FNC}}}
\newcommand{\ESS}{\prob{Exact Secure Set}}
\newcommand{\ESSF}{\prob{Exact Secure Set\textsuperscript{F}}}
\newcommand{\ESSFN}{\prob{Exact Secure Set\textsuperscript{FN}}}
\newcommand{\ESSFNC}{\prob{Exact Secure Set\textsuperscript{FNC}}}
\newcommand{\MMO}{\prob{Minimum Maximum Outdegree}}
\newcommand{\SSV}{\prob{Secure Set Verification}}

\newcommand{\sFNC}[1]{\ensuremath{\sigma_{#1}^\text{FNC}}}
\newcommand{\tFNC}{\ensuremath{\tau^\text{FNC}}}
\newcommand{\sFN}[1]{\ensuremath{\sigma_{#1}^\text{FN}}}
\newcommand{\tFN}{\ensuremath{\tau^\text{FN}}}
\newcommand{\tF}{\ensuremath{\tau^\text{F}}}
\newcommand{\size}[1]{\ensuremath{{\lvert #1 \rvert}}}

\newcommand{\complexityClass}[1]{\ensuremath{\mathrm{#1}}}

\newcommand{\NP}{\complexityClass{NP}}
\newcommand{\CONP}{\complexityClass{\textrm{co-}NP}}
\newcommand{\Sptwo}{\complexityClass{\Sigma^P_2}}
\newcommand{\FPT}{\complexityClass{FPT}}
\newcommand{\Wone}{\complexityClass{W[1]}}
\newcommand{\XP}{\complexityClass{XP}}
\newcommand{\T}{\ensuremath{\mathcal{T}}}

\newcommand{\NX}{\ensuremath{\overline{X}}}
\newcommand{\NY}{\ensuremath{\overline{Y}}}
\newcommand{\NT}{\ensuremath{\overline{T}}}
\newcommand{\NTP}{\ensuremath{\overline{T'}}}
\newcommand{\NTS}{\ensuremath{\overline{T}_\square}}
\newcommand{\YT}{\ensuremath{Y_\triangle}}
\newcommand{\YPT}{\ensuremath{Y'_\triangle}}
\newcommand{\YS}{\ensuremath{Y_\square}}
\newcommand{\NTT}{\ensuremath{\overline{T}_\triangle}}
\newcommand{\TP}{\ensuremath{T'}}
\newcommand{\TPS}{\ensuremath{T'_\square}}
\newcommand{\NTPS}{\ensuremath{\overline{T'}_\square}}

\newcommand{\true}{\ensuremath{\text{true}}}
\newcommand{\false}{\ensuremath{\text{false}}}

\newcommand{\hatS}{\ensuremath{\widehat{S}}}
\newcommand{\hatSp}{\ensuremath{\widehat{S'}}}
\newcommand{\hatSpp}{\ensuremath{\widehat{S''}}}
\newcommand{\hatX}{\ensuremath{\widehat{X}}}
\newcommand{\hatXp}{\ensuremath{\widehat{X'}}}
\newcommand{\hatXpp}{\ensuremath{\widehat{X''}}}
\newcommand{\score}[2]{\ensuremath{\operatorname{score}_{#1,#2}}}
\newcommand{\scorehatSt}{\ensuremath{\score{\hatS}{t}}}
\newcommand{\scorehatStp}{\ensuremath{\score{\hatS}{t'}}}
\newcommand{\scorehatStpp}{\ensuremath{\score{\hatS}{t''}}}
\newcommand{\scorehatSthatX}{\ensuremath{\scorehatSt(\hatX)}}
\newcommand{\scorehatSthatXp}{\ensuremath{\scorehatSt(\hatXp)}}
\newcommand{\scorehatSthatXpp}{\ensuremath{\scorehatSt(\hatXpp)}}
\newcommand{\scorehatStphatX}{\ensuremath{\scorehatStp(\hatX)}}
\newcommand{\scorehatStphatXp}{\ensuremath{\scorehatStp(\hatXp)}}
\newcommand{\scorehatStpphatXpp}{\ensuremath{\scorehatStpp(\hatXpp)}}
\newcommand{\lscore}[2]{\ensuremath{\operatorname{lscore}_{#1,#2}}}
\newcommand{\lscoreSt}{\ensuremath{\lscore{S}{t}}}
\newcommand{\lscoreStX}{\ensuremath{\lscoreSt(X)}}
\newcommand{\lscorehatSt}{\ensuremath{\lscore{\hatS}{t}}}
\newcommand{\lscorehatStX}{\ensuremath{\lscorehatSt(X)}}
\newcommand{\cval}[2]{\ensuremath{c_{#1,#2}}}
\newcommand{\cvalhatSt}{\ensuremath{\cval{\hatS}{t}}}
\newcommand{\cvalhatStp}{\ensuremath{\cval{\hatS}{t'}}}
\newcommand{\cvalhatStpp}{\ensuremath{\cval{\hatS}{t''}}}
\newcommand{\cvalhatStX}{\ensuremath{\cvalhatSt(X)}}
\newcommand{\cvalhatStpX}{\ensuremath{\cvalhatStp(X)}}
\newcommand{\cvalhatStpXp}{\ensuremath{\cvalhatStp(X')}}
\newcommand{\cvalhatStppX}{\ensuremath{\cvalhatStpp(X)}}
\newcommand{\hatSt}{\ensuremath{\hatS_t}}
\newcommand{\hatStp}{\ensuremath{\hatS_{t'}}}
\newcommand{\hatStpp}{\ensuremath{\hatS_{t''}}}
\newcommand{\tables}[2]{\ensuremath{C_{#1,#2}}}
\newcommand{\tablesSt}{\ensuremath{\tables{S}{t}}}
\newcommand{\tablesStp}{\ensuremath{\tables{S}{t'}}}
\newcommand{\tablesStpp}{\ensuremath{\tables{S}{t''}}}
\newcommand{\possibleTables}[1]{\ensuremath{F_{#1}}}
\newcommand{\possibleTablesS}{\ensuremath{\possibleTables{S}}}
\newcommand{\possibleTablesSsetminusv}{\ensuremath{\possibleTables{S \setminus \{v\}}}}
\newcommand{\possibleTablesScupv}{\ensuremath{\possibleTables{S \cup \{v\}}}}

\newcommand{\origin}{\ensuremath{\operatorname{origin}_{S,t}}}
\newcommand{\originS}{\ensuremath{\operatorname{origin}_{S,t}^{\in}}}
\newcommand{\originSc}{\ensuremath{\originS(c)}}
\newcommand{\originNotS}{\ensuremath{\operatorname{origin}_{S,t}^{\notin}}}
\newcommand{\originNotSc}{\ensuremath{\originNotS(c)}}
\newcommand{\originJoinSt}{\ensuremath{\operatorname{origin}_{S,t}}}
\newcommand{\originJoinStc}{\ensuremath{\originJoinSt(c)}}

\newcommand{\plusS}{\ensuremath{\oplus_{S,t}}}
\newcommand{\plusScupv}{\ensuremath{\oplus_{S \cup \{v\},t}}}
\newcommand{\minusS}{\ensuremath{\ominus_{S,t}^{\in}}}
\newcommand{\minusNotS}{\ensuremath{\ominus_{S,t}^{\notin}}}
\newcommand{\joinSt}{\ensuremath{\otimes_{S,t}}}

\title{Complexity of Secure Sets\footnote{This work was supported by the Austrian Science Fund (FWF) projects P25607 and Y698.
The authors would like to thank the anonymous referees of a journal submission
of this paper for their valuable feedback.}}

\makeatletter
\@ifclassloaded{lipics}{
\theoremstyle{plain}
\newtheorem{observation}[theorem]{Observation}

\author{Bernhard Bliem}
\author{Stefan Woltran}
\affil{%
Institute of Information Systems 184/2\\
TU Wien\\
Favoritenstrasse 9--11, 1040 Vienna, Austria\\
\texttt{[bliem,woltran]@dbai.tuwien.ac.at}%
}
\authorrunning{B. Bliem and S. Woltran} 

\Copyright{Bernhard Bliem and Stefan Woltran}


}{
\usepackage[a4paper,margin=3.5cm]{geometry}
\usepackage{amssymb}
\usepackage[tbtags,fleqn]{amsmath}
\usepackage{amsthm}
\theoremstyle{plain}
\newtheorem{theorem}{Theorem}
\newtheorem{lemma}[theorem]{Lemma}
\newtheorem{corollary}[theorem]{Corollary}
\newtheorem{observation}[theorem]{Observation}
\theoremstyle{definition}
\newtheorem{definition}[theorem]{Definition}

\theoremstyle{remark}

\ifx\numberwithinsect\relax
  \@addtoreset{theorem}{section}
  \edef\thetheorem{\expandafter\noexpand\thesection\@thmcountersep\@thmcounter{theorem}}
\fi

\ifnum\pdfshellescape=1
  \usetikzlibrary{external}
  \immediate\write18{mkdir -p out/out}
  \immediate\write18{touch out/out/directory-exists}
  \IfFileExists{out/out/directory-exists}{
    \tikzsetexternalprefix{out/}
  }{}
  \tikzexternalize
\fi

\author{Bernhard Bliem and Stefan Woltran}
}
\makeatother

\begin{document}

\maketitle

\begin{abstract}
A secure set $S$ in a graph is defined as a set of vertices such that for any
$X\subseteq S$  the majority of vertices in the neighborhood of $X$ belongs to
$S$.
It is known that deciding whether a set $S$ is secure in a graph is
$\CONP$-complete.
However, it is still open how this result contributes to the actual complexity
of deciding whether for a given graph $G$ and integer $k$, a non-empty secure
set for $G$ of size at most $k$  exists.
In this work, we pinpoint the complexity of this problem by showing that it is
\Sptwo{}-complete.
Furthermore, the problem has so far not been subject to a parameterized
complexity analysis that considers structural parameters.
In the present work, we prove that the problem is \Wone{}-hard when
parameterized by treewidth.
This is surprising since the problem is known to be FPT when parameterized by
solution size and ``subset problems'' that satisfy this property usually tend
to be FPT for bounded treewidth as well.
Finally, we give an upper bound by showing membership in \XP{}, and we provide
a positive result in the form of an FPT algorithm for checking whether a given
set is secure on graphs of bounded treewidth.
\end{abstract}

\section{Introduction}

The objective of many problems that can be modeled as graphs is finding a group
of vertices that together satisfy some property.
In this respect, one of the concepts that has been quite extensively
studied~\cite{yero2013defensive} is the notion of a defensive
alliance~\cite{DBLP:journals/combinatorics/HaynesHH03}, which is a set of
vertices such that for each element $v$ at least half of its neighbors are also
in the alliance.
The name ``defensive alliance'' stems from the intuition that the neighbors of
such an element $v$ that are also in the alliance can help out in case $v$ is
attacked by its other neighbors.
Notions like this can be applied to finding groups of nations, companies or
individuals that depend on each other, but also to more abstract situations
like finding groups of websites that form
communities~\cite{DBLP:journals/computer/FlakeLGC02}.

In this work, we are looking at a natural generalization of defensive alliances
called \emph{secure sets}, which have been introduced by Brigham
et~al.~\cite{DBLP:journals/dam/BrighamDH07}.
While defensive alliances make sure that each element of an alliance can defend
itself against attacks from its neighbors, they do not account for attacks on
multiple vertices at the same time.
To this end, we can employ a stronger concept:
A secure set of a graph $G$ is a subset $S$ of the vertices of $G$ such that
for each $X\subseteq S$, the number of vertices in $N[X]\cap S$ is not less
than the number of vertices in $N[X]\setminus S$.
Here $N[X]$ denotes the closed neighborhood of $X$ in $G$, i.e., $X$ together
with all vertices adjacent to $X$.
The \probSS{} problem can now be stated as follows:
Given a graph $G$ and an integer $k$, does there exists a secure set $S$
of $G$ such that
$1\leq \size{S} \leq k$? 

It is known that deciding whether a \emph{given} set $S$ is secure in a graph
is $\CONP$-complete~\cite{Ho2011}, so it would not be surprising if
\emph{finding} (non-trivial) secure sets is also a very hard problem.
Unfortunately, the exact complexity of 
this problem
has so far remained unresolved.
This is an unsatisfactory state of affairs because it leaves the possibility open that existing approaches for solving the problem (e.g., \cite{jlc:AbseherBCDW15}) are suboptimal in that they employ unnecessarily powerful programming techniques.
Hence we require a precise complexity-theoretic classification of the problem.

Due to its high complexity, it makes sense to look at the parameterized
complexity~\cite{downey1999parameterized,flum2006parameterized,niedermeier2006invitation,Cygan15}
of the problem and to study if \probSS{} becomes tractable under the assumption
that certain parameters of the problem instances are small.
For some parameters, this may be a reasonable assumption in practice.
For instance, it has been shown that \probSS{} can be solved in linear time if
the solution size is bounded by a constant~\cite{DuttonE08}.
If we are only interested in small secure sets, the resulting algorithm is
therefore a good choice.

However, we often cannot make the assumption that the solutions are small.
In such cases, it is a common strategy to consider structural parameters
instead, which measure in a certain way how complex the graph underlying a
problem instance is.
One of the most studied structural parameters is
treewidth~\cite{DBLP:journals/jct/RobertsonS84,DBLP:journals/actaC/Bodlaender93,DBLP:conf/sofsem/Bodlaender05},
which indicates how close a graph is to being a tree.
Treewidth is an attractive parameter because many hard problems become
tractable on instances of bounded treewidth, and in several practical
applications it has been observed that the considered problem instances exhibit
small
treewidth~\cite{DBLP:journals/actaC/Bodlaender93,iandc:Thorup98,dam:KornaiT92}.
%
In~\cite{ho2009rooted} it has been shown that a certain variant of \probSS{}
becomes easy on trees, but the complexity of \probSS{} parameterized by
treewidth is listed as an open problem in that work and has so far remained
unresolved.

The first main contribution of our paper is to show that 
\probSS{} 
is $\Sptwo$-complete. Unlike the existing
$\CONP$-hardness proof~\cite{Ho2011},
which uses a (quite involved) 
reduction from \prob{Dominating Set}, we base our proof on a reduction from a 
problem in the area of logic. To be specific, we first show that the canonical
$\Sptwo$-complete problem \prob{Qsat$_2$} can be reduced to a 
variant of \probSS{}, where
vertices can be forced
to be in or out of every solution,
and pairs of vertices can be 
specified to indicate that every solution must contain exactly one element of each such pair.
In order to prove the desired complexity result,
we then successively reduce this variant to the standard 
\probSS{} problem. 
At the same time, we show \Sptwo{}-completeness for the exact variants of these
problems, where we are interested in secure sets \emph{exactly} of a certain
size.

Membership in the class $\Sptwo$ is rather obvious;
in fact, 
\cite{jlc:AbseherBCDW15} presents a polynomial-time 
reduction to Answer Set Programming \cite{DBLP:journals/cacm/BrewkaET11} and thus shows this result implicitly.
Together with our corresponding hardness result, it follows
that 
\probSS{} is $\Sptwo$-complete, and it turns out
that
all the problem variants we consider in this paper
are $\Sptwo$-complete.

We thus complete the picture of the precise complexity of the \probSS{}
problem, and we also provide completeness results for variants of the problem
that have already been proposed~\cite{ho2009rooted} but for which no complexity
analysis has been performed so far.
Our results underline that
\probSS{} is among the few rather natural problems in graph theory
that are complete for the second layer of the polynomial hierarchy
(like, e.g., \prob{Clique Coloring} \cite{DBLP:journals/tcs/Marx11} or
\prob{2-Coloring Extension} \cite{DBLP:journals/amai/Szeider05}).
Moreover, $\Sptwo$-hardness of \probSS{} indicates that 
an efficient reduction to the \prob{Sat} problem is not possible (unless the
polynomial hierarchy collapses).

The second main contribution of our paper is a parameterized complexity
analysis of \probSS{} with treewidth as the parameter.
We show that this problem is hard for the class \Wone{}, which rules out a
fixed-parameter tractable algorithm under commonly held complexity-theoretic
assumptions.
This result is rather surprising for two reasons:
First, the problem is tractable on trees~\cite{ho2009rooted} and quite often
problems that become easy on trees turn out to become easy on graphs of bounded
treewidth.%
\footnote{To be precise, \cite{ho2009rooted} shows that a slight variant of
\probSS{} is tractable on trees, since \probSS{} on trees is trivial. Our
results, however, also imply \Wone{}-hardness for this particular variant.}
Second, this makes \probSS{} one of the very few ``subset problems'' that are
fixed-parameter tractable w.r.t.\ solution size but not w.r.t.\ treewidth.
Problems with this kind of behavior are rather rare, as observed
by Dom et~al.~\cite{DBLP:conf/iwpec/DomLSV08}.

Beside this parameterized hardness result, we also give an upper bound by
showing that \probSS{} is in the class \XP{}, which means that it can be solved
in polynomial time on instances of bounded treewidth.
We do so by providing an algorithm where the degree of the polynomial depends
on the treewidth.

Finally, we present a positive result for the \CONP{}-complete problem of
checking whether a given set of vertices is secure in a graph:
We provide an algorithm that solves the problem in linear time for graphs of
bounded treewidth.

This paper is organized as follows:
We first provide the necessary background in Section~\ref{sec:background}.
Then we analyze the complexity of \probSS{} in
Section~\ref{sec:classical-complexity}, where we show that this problem,
along with several variants, is \Sptwo{}-complete.
In Section~\ref{sec:parameterized-complexity-treewidth}, we consider the
parameterized complexity of \probSS{} where treewidth is our parameter of
interest.
Section~\ref{sec:conclusion} concludes the paper with a discussion.

The present work extends a conference paper~\cite{DBLP:conf/wg/BliemW15}, which
did not contain any results about the parameterized complexity of the
considered problems.
Beside the new results in Section~\ref{sec:parameterized-complexity-treewidth},
we also slightly modified some of the reductions that prove
\Sptwo{}-hardness so that they preserve bounded treewidth, which allows us to
reuse them for our parameterized hardness proofs.
We also added a reduction
(which eliminates necessary vertices),
which made one of the reductions
(from the exact variant of the problem to the non-exact variant)
from the previous paper redundant.

\section{Background}
\label{sec:background}

All graphs are undirected and simple unless stated otherwise.
We denote the set of vertices and edges of a graph $G$ by $V(G)$ and $E(G)$,
respectively.
We denote an undirected edge between vertices $u$ and $v$ as $(u,v)$ or
equivalently $(v,u)$.
It will be clear from the context whether an edge $(u,v)$ is directed or
undirected.
Given a graph $G$, the \emph{open neighborhood} of a vertex $v \in V(G)$,
denoted by $N_G(v)$, is the set of all vertices adjacent to $v$, and
$N_G[v] = N_G(v) \cup \{v\}$ is called the \emph{closed neighborhood} of $v$.
Let $S \subseteq V(G)$. We abuse notation by writing 
$N_G(S)$
and $N_G[S]$ to denote $\bigcup_{v \in S} N_G(v)$ and
$\bigcup_{v \in S} N_G[v]$, respectively.
If it is clear from the context which graph is meant, we write $N(\cdot)$ and
$N[\cdot]$ instead of $N_G(\cdot)$ and $N_G[\cdot]$, respectively.


\begin{definition}
Given a graph $G$, a set $S \subseteq V(G)$ is \emph{secure in $G$} if for each 
$X \subseteq S$ it holds that
$\size{N[X] \cap S} \geq \size{N[X] \setminus S}$.
\end{definition}
We often write ``$S$ is secure'' instead of ``$S$ is secure in $G$'' if it is clear from the context which graph is meant.
By definition, the empty set is secure in any graph. Thus, 
in the following decision problems we ask for secure sets of size at least~1.
The following is our main problem:
\decisionProblem{\probSS}{A graph $G$ and an integer $k$ with $1 \leq k \leq \size{V(G)}$}{%
Does there exist a set $S \subseteq V(G)$ with $1 \leq \size{S} \leq k$ that is secure?%
}
\input{secure-set-example-figure}
Figure~\ref{fig:secure-set-example} shows a graph together with a minimum non-empty secure set $S = \{a,b,c\}$.
Observe that for any $X \subseteq S$ the condition $\size{N[X] \cap S} \geq \size{N[X] \setminus S}$ is satisfied.

Note that the well-known \prob{Defensive Alliance} problem is a special case of \probSS{} where only those subsets $X$ of $S$ are considered that have size~$1$.
For example, in Figure~\ref{fig:secure-set-example}, the set $S' = \{a,b\}$ is a defensive alliance as $\size{N[v] \cap S'} \geq \size{N[v] \setminus S'}$ holds for each $v \in S'$, but $S'$ is not a secure set, since for $X' = S'$ it holds that $\size{N[X'] \cap S'} < \size{N[X'] \setminus S'}$.

We now define three variants of the \probSS{} problem that we require in our proofs.
\SSF{} generalizes the \probSS{} problem by designating some ``forbidden'' vertices that may never be in any solution.
This variant can be formalized as follows:
\decisionProblem{\SSF}{A graph $G$, an integer $k$ and a set $V_\square \subseteq V(G)$}{%
Does there exist a set $S \subseteq V(G) \setminus V_\square$ with $1 \leq \size{S} \leq k$ that is secure?%
}
\SSFN{} is a further generalization that, in addition, allows ``necessary'' vertices to be specified that must occur in every solution.
\decisionProblem{\SSFN}{A graph $G$, an integer $k$, a set $V_\square \subseteq V(G)$ and a set $V_\triangle \subseteq V(G)$}{%
Does there exist a set $S \subseteq V(G) \setminus V_\square$ with $V_\triangle \subseteq S$ and $1 \leq \size{S} \leq k$ that is secure?%
}
Finally, we introduce the generalization \SSFNC{}.
Here we may state pairs of ``complementary'' vertices where each solution must contain exactly one element of every such pair.
\decisionProblem{\SSFNC}{A graph $G$, an integer $k$, a set $V_\square \subseteq V(G)$, a set $V_\triangle \subseteq V(G)$ and a set $C \subseteq V(G)^2$}{%
Does there exist a set $S \subseteq V(G) \setminus V_\square$ with $V_\triangle \subseteq S$ and $1 \leq \size{S} \leq k$ that is secure and, for each pair $(a,b) \in C$, contains either $a$ or $b$?%
}
For our results on structural parameters, we need a way to represent the 
structure of a \SSFNC{} instance by a graph that augments $G$ with the 
information in $C$:
\begin{definition}
Let $I$ be a \SSFNC\ instance, let $G$ be the graph in $I$ and let $C$ the set 
of complementary vertex pairs in $I$.
By the \emph{primal graph} of $I$ we mean the graph $G'$ with
$V(G') = V(G)$ and
$E(G') = E(G) \cup C$.
\end{definition}
While the \probSS{} problem asks for secure sets of size \emph{at most} $k$, we also consider the \ESS{} problem that concerns secure sets of size \emph{exactly} $k$.
Note that a secure set may become insecure by adding or removing elements, so this is a non-trivial problem variant.
Analogously, we also define exact versions of the three generalizations of \probSS{} presented above.

When the task is not to find secure sets but to verify whether a given set is
secure, the following problem is of interest:
\decisionProblem{\SSV}{A graph $G$ and a set $S \subseteq V(G)$}{%
Is $S$ secure?%
}
This problem is known to be \CONP{}-complete~\cite{Ho2011}.

In this paper's figures, we often indicate necessary vertices by means of a triangular node shape, and forbidden vertices by means of either a square node shape or a superscript square in the node name.
If two vertices are complementary, we often express this in the figures by putting a $\neq$ sign between them.
\input{secure-set-example-figure2}
For example, in Figure~\ref{fig:secure-set-example2}, the vertices $b$ and $c$ are complementary and occur in no solution together; also the vertices $b$ and $e$ are complementary.
Note, however, that by putting a $\neq$ sign between two vertices we do not mean to express that there is an edge between them.
For instance, there is no edge between $b$ and $c$, but there is an edge between $b$ and $e$, which is explicitly drawn.
The vertex $a$ and the ``anonymous'' vertex adjacent to $c$ are necessary and occur in every solution; $d^\square$ and the ``anonymous'' vertex adjacent to $e$ are forbidden and occur in no solution.
In this figure, the unique minimum non-empty secure set satisfying the conditions of forbidden, necessary and complementary vertices consists of $a$, $b$ and the ``anonymous'' necessary vertex adjacent to $c$.

The following terminology will be helpful:
We often use the terms \emph{attackers} and \emph{defenders} of a subset $X$ of a secure set candidate $S$.
By these we mean the sets $N[X] \setminus S$ and $N[X] \cap S$, respectively.
To show that a subset $X$ of a secure set candidate $S$ is \emph{not} a witness
to $S$ being insecure, we sometimes employ the notion of a \emph{defense} of
$X$ w.r.t.\ $S$, which assigns to each attacker a dedicated defender:
If we are able to find an injective mapping $\mu: N[X] \setminus S \to N[X] \cap S$, then obviously $\size{N[X] \setminus S} \leq \size{N[X] \cap S}$, and we call $\mu$ a \emph{defense} of $X$ w.r.t.\ $S$.
Given such a defense $\mu$, we say that a defender $d$ \emph{repels} an attack on $X$ by an attacker $a$ whenever $\mu(a) = d$.
Consequentially, when we say that a set of defenders $D$ \emph{can repel} attacks on $X$ from a set of attackers $A$, we mean that there is a defense that assigns to each element of $A$ a dedicated defender in $D$.

To warm up, we make some easy observations that we will use in our proofs.
First, for every set $R$ consisting of a majority of neighbors of a vertex $v$,
whenever $v$ is in a secure set, also some element of $R$ must be in it:
\begin{observation}
\label{obs:one-in-s}
Let $S$ be a secure set in a graph,
let $v \in S$ and
let $R \subseteq N(v)$.
If $\size{R} > \frac{1}{2} N[v]$,
then $S$ contains an element of $R$.
\end{observation}
\begin{proof}
Suppose that $\size{R} > \frac{1}{2} \size{N[v]}$ and
$S$ contains no element of $R$.
Since all elements of $R$ attack $v$,
$\size{N[v] \setminus S} > \frac{1}{2} \size{N[v]}$.
Hence
$2 \size{N[v] \setminus S} > \size{N[v]} =
\size{N[v] \cap S} + \size{N[v] \setminus S}$, and we obtain
the contradiction
$\size{N[v] \setminus S} > \size{N[v] \cap S}$.
\end{proof}
Next, if one half of the neighbors of an element $v$ of a secure set attacks
$v$, then the other half of the neighbors must be in the secure set:
\begin{observation}
\label{obs:all-in-s}
Let $S$ be a secure set in a graph,
let $v \in S$ and
let $N(v)$ be partitioned into two equal-sized sets $A,D$.
If $A \cap S = \emptyset$, then $D \subseteq S$.
\end{observation}
\begin{proof}
Since $N(v)$ is partitioned into $A$ and $D$ such that
$A \cap S = \emptyset$,
we get
$N(v) \cap S = D \cap S$.
If some element of $D$ is not in $S$, then
$D \cap S \subset D$ and
$A \subset N[v] \setminus S$.
By $\size{D} = \size{A}$, we get
$\size{D \cap S} + 2 \leq \size{N[v] \setminus S}$.
From
$\size{N[v] \cap S} = 1 + \size{N(v) \cap S} = 1 + \size{D \cap S}$
we now obtain the contradiction
$\size{N[v] \cap S} < \size{N[v] \setminus S}$.
\end{proof}
In particular, if half of the neighbors of $v$ are forbidden, then $v$ can only
be in a given secure set if all non-forbidden neighbors are also in the secure set.

Finally, we recapitulate some background from complexity theory.
The class \Sptwo{} is the class of problems that are solvable in polynomial 
time by a nondeterministic Turing machine that has access to an \NP{} oracle.
The canonical problem complete for this class is \prob{Qsat$_2$}, which asks,
given a formula
$\exists x_1 \dots \exists x_{n_x} \forall y_1 \dots \forall y_{n_y} \psi$,
where $\psi$ is a propositional $3$-DNF formula,
whether there is a truth assignment to the $x_i$ variables such that for all 
truth assignments to the $y_i$ variables $\psi$ evaluates to true.

In parameterized complexity 
theory~\cite{downey1999parameterized,flum2006parameterized,niedermeier2006invitation,Cygan15}, 
we study problems that consist not only of an input and a question, but also of 
some parameter of the input that is represented as an integer.
A problem is in the class \FPT{} (``fixed-parameter tractable'') if it can be 
solved in time $f(k) \cdot n^c$, where $n$ is the input size, $k$ is the 
parameter, $f$ is a computable function that only depends on $k$, and $c$ is a 
constant that does not depend on $k$ or $n$.
We call such an algorithm an \emph{FPT algorithm}, and we call it
\emph{fixed-parameter linear} if $c=1$.
Similarly, a problem is in the class \XP{} (``slice-wise polynomial'') if it 
can be solved in time $f(k) \cdot n^{g(k)}$, where $f$ and $g$ are computable 
functions.
Note that here the degree of the polynomial may depend on $k$, so such 
algorithms are generally slower than FPT algorithms.
For the class \Wone{} it holds that $\FPT \subseteq \Wone \subseteq \XP$,
and it is commonly believed that the inclusions are proper, i.e.,
\Wone{}-hard problems do not admit FPT algorithms.
\Wone{}-hardness of a problem can be shown using parameterized reductions, 
which are reductions that run in FPT time and produce an equivalent instance
whose parameter is bounded by a function of the original parameter.

For problems whose input can be represented as a graph, one important parameter 
is \emph{treewidth}, which is a structural parameter that, roughly speaking, 
measures the ``tree-likeness'' of a graph. It is defined by means of tree 
decompositions, originally introduced in~\cite{DBLP:journals/jct/RobertsonS84}.
The intuition behind tree decompositions is to obtain a tree from a 
(potentially cyclic) graph by subsuming multiple vertices under one node and 
thereby isolating the parts responsible for cyclicity.
\begin{definition}
\label{def:td}
A \emph{tree decomposition} of a graph $G$
is a pair $\T = (T,\chi)$ where $T$ is a (rooted) tree and
$\chi : V(T) \to 2^{V(G)}$ assigns to each node of $T$ a set of vertices of $G$
(called the node's \emph{bag}), such that the following conditions are met:
\begin{enumerate}
\item For every vertex $v \in V(G)$, there is a node $t \in V(T)$ such that $v 
\in \chi(t)$.
\item For every edge $(u,v) \in E(G)$, there is a node $t \in V(T)$ such that
$\{u,v\} \subseteq \chi(t)$.
\item For every $v \in V(G)$, the subtree of $T$ induced by $\{t \in V(T) \mid 
v \in \chi(t)\}$ is connected.
\end{enumerate}
We call $\max_{t \in V(T)} \lvert \chi(t) \rvert - 1$ the \emph{width} of $\T$.
The \emph{treewidth} of a graph is the minimum width over all its tree 
decompositions.
\end{definition}
In general, constructing an optimal tree decomposition (i.e., a tree
decomposition with minimum width) is
intractable~\cite{Arnborg:1987:CFE:37170.37183}.
However, the problem is solvable in linear time on graphs of bounded treewidth
(specifically in time
$w^{\O(w^3)} \cdot n$, where $w$ is the
treewidth)~\cite{DBLP:journals/siamcomp/Bodlaender96} and
there are also heuristics that offer good performance in
practice~\cite{DBLP:conf/micai/DermakuGGMMS08,DBLP:journals/iandc/BodlaenderK10}.

In this paper we will consider so-called \emph{nice} tree decompositions:
\begin{definition}
\label{def:nice-td}
A tree decomposition $\mathcal{T} = (T,\chi)$ is
\emph{nice} if each node $t \in V(T)$ is of one of the following types:
\begin{enumerate}
 \item Leaf node: The node $t$ has no child nodes.
 \item Introduce node: The node $t$ has exactly one child node $t'$ such that $\chi(t) \setminus \chi(t')$ consists of exactly one element.
 \item Forget node: The node $t$ has exactly one child node $t'$ such that $\chi(t') \setminus \chi(t)$ consists of exactly one element.
 \item Join node: The node $t$ has exactly two child nodes $t_1$ and $t_2$ with $\chi(t) = \chi(t_1) = \chi(t_2)$.
\end{enumerate}
Additionally, the bags of the root and the leaves of $T$ are empty.
\end{definition}
A tree decomposition of width $w$ for a graph with $n$ vertices can be
transformed into a nice one of width $w$ with $\O(wn)$ nodes in fixed-parameter
linear time~\cite{Kloks1994treewidth}.

\input{td-example-figure}

For any tree decomposition $\T$ and an element $v$ of some bag in $\T$,
we use the notation $t_v^\T$ to denote the unique ``topmost node'' whose bag
contains $v$ (i.e., $t_v^\T$ does not have a parent whose bag contains $v$).
Figure~\ref{fig:td-example} depicts a graph and a nice tree decomposition,
where we also illustrate the $t^\T_v$ notation.

When we speak of the treewidth of an instance of \probSS{}, \SSF{}, \SSFN{}, \ESS{}, \ESSF{} or
\ESSFN{}, we mean the treewidth of the graph in the instance.
For an instance of \SSFNC{} or \ESSFNC{}, we mean the treewidth of the primal graph.

\section{Complexity of the Secure Set Problem}
\label{sec:classical-complexity}

This section is devoted to proving the following theorem:
\begin{theorem}
\label{thm:ss-variants-sptwo}
The following problems are all $\Sptwo$-complete:
\probSS{},
\ESS{},
\SSF{},
\ESSF{},
\SSFN{},
\ESSFN{},
\SSFNC{} and
\ESSFNC{}.
\end{theorem}
We prove this by providing a chain of polynomial reductions from
\prob{Qsat$_2$} to the problems under consideration.

\subsection{Hardness of Secure Set With Forbidden, Necessary And Complementary Vertices}

\begin{lemma}
\SSFNC{} and \ESSFNC{} are \Sptwo-hard.
\label{lem:qsat-to-ssfnc-correct}
\end{lemma}

\begin{proof}
We reduce from \prob{Qsat$_2$} to \SSFNC{}.
This also proves \Sptwo-hardness for the exact variant because our reduction
makes sure that all solutions of the \SSFNC{} instance have the same size.
We are given a quantified Boolean formula
$\phi = \exists x_1 \dots \exists x_{n_x} \forall y_1 \dots \forall y_{n_y} \psi$,
where $\psi$ is in $3$-DNF and contains $n_t$ terms.
We assume that no term contains both a variable and its complement (since such
a term can never be satisfied) and that each term contains at least one
universally quantified variable (since $\phi$ is trivially true otherwise).

We construct an instance $(G,k,V_\triangle,V_\square,C)$ of \SSFNC{} in the following.
\input{qsat-reduction-figure}
For an illustration, see Figure~\ref{fig:qsat-reduction-example}.
We define a graph $G$ by choosing the union of the following sets as $V(G)$:
\begingroup
\allowdisplaybreaks
\begin{align*}
X ={}& \{x_1, \dots, x_{n_x} \} &
\NX ={}& \{\overline{x_1}, \dots, \overline{x_{n_x}} \} \\
Y ={}& \{y_1, \dots, y_{n_y} \} &
\NY ={}& \{\overline{y_1}, \dots, \overline{y_{n_y}} \} \\
\YT ={}& \{y_{i,j}^\triangle, \overline{y_{i,j}}^\triangle \mid 1 \leq i \leq n_y,\; 1 \leq j \leq n_t \} &
\YPT ={} &\{y_j^\triangle \mid 1 \leq j \leq n_t - 1 \} \\
\YS ={}& \{y_{i,j}^\square \mid 1 \leq i \leq n_y,\; 1 \leq j \leq n_t+1 \} &
H ={}& \{d_1^\square, d_2^\square, \overline{t}^\square\}\\
T ={}& \{t_1, \dots, t_{n_t}\} &
\NT ={}& \{\overline{t_1}, \dots, \overline{t_{n_t}} \} \\
\NTS ={}& \{\overline{t_1}^\square, \dots, \overline{t_{n_t}}^\square\} &
\NTT ={}& \{\overline{t_1}^\triangle, \dots, \overline{t_{n_t}}^\triangle \} \\
\TP ={}& \{t_1', \dots, t_{n_t}'\} &
\NTP ={}& \{\overline{t'_1}, \dots, \overline{t'_{n_t}} \} \\
\TPS ={}& \{t_1'^\square, \dots, t_{n_t}'^\square\} &
\NTPS ={}& \{\overline{t_1'}^\square, \dots, \overline{t_{n_t}'}^\square\}
\end{align*}
\endgroup
Next we define the set of edges.
In the following, whenever we sloppily speak of a literal in the context of the
graph $G$, we mean the vertex corresponding to that literal (i.e., some $x_i$,
$\overline{x_i}$, $y_i$ or $\overline{y_i}$), and we proceed similarly for
terms.
Furthermore, when we are dealing with a (vertex corresponding to a) literal
$l$, then $\overline{l}$ shall denote the (vertex corresponding to the)
complement of $l$.
For any term $t_i$, let $L_X(t_i)$ and $L_Y(t_i)$ denote the set of
existentially and universally quantified literals, respectively, in $t_i$.
\begingroup
\allowdisplaybreaks
\begin{align*}
E(G) ={} &\left\{(\overline{t_i},\overline{t}^\square), (\overline{t_i},\overline{t_i}^\triangle), (t_i',t_i'^\square), (\overline{t_i'},\overline{t_i'}^\square) \mid t_i \in T\right\} \cup \left(\TP \times (Y \cup \NY)\right)\\
{}\cup{} & \left\{(\overline{l},\overline{t_i}^\square), (\overline{l},\overline{t_i}) \mid t_i \in T,\; l \in L_X(t_i)\right\}
\cup \left\{(\overline{l},\overline{t_i'}) \mid t_i \in T,\; l \in L_Y(t_i)\right\}\\
{}\cup{} & \left\{(d_1^\square,\overline{t_i}) \mid t_i \in T,\; \size{L_X(t_i)} \leq 1\right\}
\cup \left\{(d_2^\square,\overline{t_i}) \mid t_i \in T,\; L_X(t_i) = \emptyset\right\}\\
{}\cup{} & \left\{(y_i,y_{i,j}^\triangle), (\overline{y_i},\overline{y_{i,j}}^\triangle) \mid 1 \leq i \leq n_y,\; 1 \leq j \leq n_t\right\}\\
{}\cup{} & \left\{(y_i,y_{i,j}^\square), (\overline{y_i},y_{i,j}^\square) \mid y_{i,j}^\square \in \YS\right\}
\cup \left(\YPT \times (Y \cup \NY)\right)
\end{align*}
\endgroup
Finally, we define
\begin{align*}
V_\triangle ={}& Y \cup \NY \cup \YT \cup \YPT \cup \NTT, \quad
V_\square = \YS \cup \NTS \cup \TPS \cup \NTPS \cup H,\\
C ={}& \{(x_i,\overline{x_i}) \mid 1 \leq i \leq n_x\} \cup \{(t_i,\overline{t_i}), (\overline{t_i},t_i'), (t_i',\overline{t_i'}) \mid 1 \leq i \leq n_t\},
\end{align*}
and $k = \size{V_\triangle} + n_x + 2n_t$.

The following observations are crucial:
Elements of $X \cup \NX$ are only adjacent to vertices from $\NTS$ (which are forbidden) and $\NT$.
For any $i$, each element of $X \cup \NX$ is adjacent to $\overline{t_i}^\square \in \NTS$ iff it is adjacent to $\overline{t_i} \in \NT$.
Furthermore, for any $i,j$, if $x_i$ or $\overline{x_i}$ is adjacent to $\overline{t_j}$, then setting the variable $x_i$ to \true\ or \false, respectively, falsifies the term $t_j$.
Finally, for any $i,j$, if $y_i$ or $\overline{y_i}$ is adjacent to $\overline{t_j'}$, then setting the variable $y_i$ to \true\ or \false, respectively, falsifies the term $t_j$.

The intuition is that the complementary pairs $(x_i,\overline{x_i})$ guess a 
truth assignment to the existentially quantified variables.
We now need to check if such a truth assignment has the property that the 
formula $\psi$ is true for all extensions of this assignment to the universally 
quantified variables.
Trying out all these extensions amounts to going through all subsets of a 
solution candidate and comparing the numbers of attackers and defenders.

To illustrate, let $S$ be a solution candidate (i.e., a set of vertices) and 
suppose $S$ satisfies the conditions on forbidden, necessary and complementary 
vertices.
We denote the truth assignment to $x_1,\dots,x_{n_x}$ encoded in $S$ by $I_S$.
Moreover, let $R$ be a subset of $S$ containing either $y_j$ or 
$\overline{y_j}$ for each universally quantified variable $y_j$.
We denote the extension of $I_S$ to $y_1,\dots,y_{n_y}$ encoded in $R$ by 
$I_{S,R}$.
For any term $t_i$ that is falsified already by $I_S$, the vertex $t_i'$ 
attacks all vertices $y_j$ and $\overline{y_j}$.
At the same time, for any term $t_i$ that is not falsified by $I_S$, the vertex 
$\overline{t_i'}$ attacks $y_j$ or $\overline{y_j}$ if setting the variable 
$y_j$ to \true\ or \false, respectively, falsifies $t_i$.
Hence, the number of attacks from vertices of the form $t_i'$ or 
$\overline{t_i'}$ on $R$ is exactly the number of terms that are falsified by 
$I_{S,R}$.
With the help of the vertices in $\YPT$, we can afford up to $n_t-1$ falsified 
terms, but if we falsify all $n_t$ terms, then $R$ is a witness that $S$ is not 
secure.

The \SSFNC{} instance $(G,k,V_\triangle,V_\square,C)$ can be constructed in
time polynomial in the size of $\phi$.
We claim that $\phi$ is true iff 
$(G,k,V_\triangle,V_\square,C)$ 
is a positive instance of \SSFNC{}.

\medskip
\noindent
\emph{``Only if'' direction.}
If $\phi$ is true, then there is an assignment $I$ to $x_1, \dots, x_{n_x}$ such that, for all assignments extending $I$ to $y_1, \dots, y_{n_y}$, some term in $\psi$ is satisfied.
We define a set
\begin{align*}
S ={} &V_\triangle
\cup \{x_i \in X \mid I(x_i) = \true\} \cup \{\overline{x_i} \in \NX \mid I(x_i) = \false\} \\
{}\cup{} &\{\overline{t_i} \in \NT,\; \overline{t_i'} \in \NTP \mid \text{there is some } l \in L_X(t_i) \text{ such that } I \not\models l\} \\
{}\cup{} &\{t_i \in T,\; t_i' \in \TP \mid \text{for all } l \in L_X(t_i) \text{ it holds that } I \models l\}.
\end{align*}
We observe that $\size{S} = k$,
$V_\square \cap S = \emptyset$,
$V_\triangle \subseteq S$, and that
for any $(a,b) \in C$ it holds that $a \in S$ iff $b \notin S$.
By construction, whenever some element of $X \cup \NX$ is in $S$, then all its
neighbors in $\NT$ are in $S$; and whenever some $\overline{t_i}$ is in $S$,
then some neighbor of $\overline{t_i}$ in $X \cup \NX$ is in $S$.

We claim that $S$ is a secure set in $G$.
Let $R$ be an arbitrary subset of $S$.
We show that $R$ has at least as many defenders as attackers by constructing a
defense, which assigns to each attacker of $R$ a dedicated defender in
$N[R] \cap S$.
We distinguish cases regarding the origins of the attacks on $R$.
\begin{itemize}
\item We repel each attacker $\overline{t_i}^\square \in \NTS$ using
$\overline{t_i}$. Since $\overline{t_i}^\square$ attacks $R$, $R$ must contain
some element of $X \cup \NX$ that is adjacent to $\overline{t_i}^\square$ and
thus also to $\overline{t_i}$, so $\overline{t_i} \in N[R] \cap S$.

\item Each attacker from $X \cup \NX \cup \{d_1^\square,d_2^\square\}$ is
adjacent to some $\overline{t_i} \in \NT \cap R$. We repel that attacker using
$\overline{t_i}^\triangle$, which is adjacent to $\overline{t_i}$. Note that it
cannot be the case that $\overline{t_i}$ is attacked by more than one vertex in
$X \cup \NX \cup \{d_1^\square,d_2^\square\}$ because $\overline{t_i}$ has
exactly two neighbors from that set and would not be in $S$ if neither of these
neighbors was in $S$.

\item If $\overline{t}^\square$ attacks $R$, then it attacks at least one
element of $\NT \cap R$, which is adjacent to some element of $X \cup \NX$ that
is also in $S$. We repel $\overline{t}^\square$ using any such element of
$X \cup \NX$.

\item Any attack from some $\overline{t_i} \in \overline{T}$ on $R$ must be on
$\overline{t_i}^\triangle$.
Since $\overline{t_i} \notin S$, $\overline{t_i}^\triangle$ is not consumed for
repelling an attack on $\overline{t_i}$, so we repel $\overline{t_i}$ with
$\overline{t_i}^\triangle$.

\item If some $t_i'^\square \in \TPS$ attacks $R$ (by attacking $t_i'$), we
repel $t_i'^\square$ with $t_i'$.

\item Analogously, we repel each attacker $\overline{t_i'}^\square \in \NTPS$
with $\overline{t_i'}$.

\item If, for some $i$ with $1 \leq i \leq n_y$, the vertices $y_{i,j}^\square$
for $1 \leq j \leq n_t+1$ attack $R$, then we distinguish the following cases:
If $y_i$ is in $R$, then the adjacent vertices $y_{i,j}^\triangle$ for
$1 \leq j \leq n_t$ are in the neighborhood of $R$, too. We then repel each
$y_{i,j}^\square$ with $y_{i,j}^\triangle$ for $1 \leq j \leq n_t$, and we
repel $y_{i,n_t+1}^\square$ with $y_i$. Otherwise, $\overline{y_i}$ is in $R$,
and we proceed symmetrically using $\overline{y_{i,j}}^\triangle$ and
$\overline{y_i}$ as dedicated defenders.

\item In order to account for attacks from $\TP \cup \NTP$ on $R$, we
distinguish two cases.
\begin{itemize}
\item If, for some $i$ with $1 \leq i \leq n_y$, both $y_i$ and
$\overline{y_i}$ are in $R$, then, in the step before, we have repelled each
$y_{i,j}^\square$ with the respective $y_{i,j}^\triangle$ or $y_i$, but all
$\overline{y_{i,j}}^\triangle$ are still free.
These vertices can repel all attacks from $\TP \cup \NTP$, as there are at most $n_t$ such attacks.

\item
Otherwise we show that there are at most $n_t - 1$ attacks from
$\TP \cup \NTP$, and they can be repelled using $\YPT$. Consider the (partial)
assignment $J$ that assigns the same values to the variables $x_1, \dots,
x_{n_x}$ as the assignment $I$ above, and, for any variable $y_i$, sets $y_i$
to $\true$ or $\false$ if $R$ contains the vertex $y_i$ or $\overline{y_i}$,
respectively. By assumption we know that our assignment to
$x_1, \dots, x_{n_x}$ is such that for all assignments to $y_1, \dots, y_{n_y}$
some term $t_i$ in $\psi$ is true. In particular, it must therefore hold that
$J$ falsifies no existentially quantified literal in $t_i$. Then, by
construction of $S$, the vertex $\overline{t_i'}$ is not in $S$.
We also know that $J$ falsifies no universally quantified literal in $t_i$. But
then the vertices from $Y \cup \NY$ adjacent to the vertex $\overline{t_i'}$
are not in $R$ due to our construction of $J$, so $\overline{t_i'}$ does not
attack any vertex in $R$. From this it follows that there are at most $n_t - 1$
attacks from $\TP \cup \NTP$ on $R$. We can repel all these attacks using the
vertices $y_1^\triangle, \dots, y_{n_t-1}^\triangle$.
\end{itemize}
\end{itemize}
This allows us to conclude $\size{N[R] \cap S} \geq \size{N[R] \setminus S}$.
Therefore $S$ is secure.

\medskip
\noindent
\emph{``If'' direction.}
Suppose $S$ is a secure set in $G$ satisfying the conditions regarding
forbidden, necessary and complementary vertices.
First observe that $\size{S} = k$ because the complementary vertex pairs make
sure that $S$ contains exactly half of
$V(G) \setminus (V_\triangle \cup V_\square)$.

If $S$ contains some $l \in X \cup \NX$, then $N(l) \cap \NT \subseteq S$
by Observation~\ref{obs:all-in-s}.
If $S$ contains some $\overline{t_i} \in \NT$, then $\overline{t_i}$ must be
adjacent to some element of $X \cup \NX$ that is also in $S$ 
by Observation~\ref{obs:one-in-s}.

We construct an interpretation $I$ on the variables $x_1, \dots, x_{n_x}$ that
sets exactly those $x_i$ to true where the corresponding vertex $x_i$ is in
$S$, and we claim that for each extension of $I$ to the universally quantified
variables there is a satisfied term in $\psi$.
To see this, suppose to the contrary that some assignment $J$ to all variables
extends $I$ but falsifies all terms in $\psi$.
Then we define a set $R$ consisting of all vertices $y_i$ such that
$J(y_i) = \true$, all vertices $\overline{y_i}$ such that $J(y_i) = \false$,
and all vertices in $(\TP \cup \NTP) \cap S$ that are adjacent to these
vertices $y_i$ or $\overline{y_i}$.
We show that this contradicts $S$ being secure:
Clearly, $R$ is a subset of $S$
and has $\size{R}$ defenders due to itself, $n_t - 1$ defenders due to
$\YPT$,
and $n_y \cdot n_t$ defenders due to 
$N(R) \cap \YT$.
This amounts to
$\size{N[R] \cap S} = \size{R} + n_t - 1 + n_y \cdot n_t$.
On the other hand, there are $n_t$ attacks on $R$ from $\TP \cup \NTP$.
This is because for any term $t_i$ in $\psi$ one of the following cases
applies:
\begin{itemize}
\item The term $t_i$ is falsified already by $I$.
Then $\overline{t_i'} \in S$ and thus $t_i' \notin S$.
The vertex $t_i'$, however, is adjacent to every element of $Y \cup \NY$, so it
attacks $R$.
\item The term $t_i$ is not falsified by $I$ but by $J$.
Then $\overline{t_i'} \notin S$, and $L_Y(t_i)$ contains some literal $l$ with
$\overline{l} \in N(\overline{t_i'})$ and $J \models \overline{l}$, so
$\overline{l}$ is in $R$ and attacked by $\overline{t_i'}$.
\end{itemize}
In addition to these $n_t$ attackers, $R$ has $\size{R \cap (\TP \cup \NTP)}$
attackers in
$N(R) \cap (\TPS \cup \NTPS)$,
as well as $n_y \cdot (n_t+1)$ attackers in
$\YS$.
As $\size{R} = n_y + \size{R \cap (\TP \cup \NTP)}$, we obtain in total
\[\size{N[R] \setminus S} =
n_t + \size{R \cap (\TP \cup \NTP)} + n_y \cdot (n_t+1) =
\size{R} + n_t + n_y \cdot n_t >
\size{N[R] \cap S}.\]
This contradicts $S$ being secure, so for each extension of $I$ to the
universally quantified vertices, $\psi$ is true; hence $\phi$ is true.
\end{proof}

\subsection{Hardness of Secure Set With Forbidden And Necessary Vertices}

Next we present a transformation $\tFNC$ that eliminates complementary vertex
pairs by turning a \SSFNC{} instance into an equivalent \SSFN{} instance.
Along with $\tFNC$, we define a function $\sFNC{I}$, for each \SSFNC{} instance
$I$,
such that the solutions of $I$ are in a one-to-one correspondence with those of
$\tFNC(I)$ in such a way that any two solutions of $I$ have the same size iff
the corresponding solutions of $\tFNC(I)$ have the same size.
We use these functions to obtain a polynomial-time reduction from \SSFNC{} to
\SSFN{} as well as from \ESSFNC{} to \ESSFN{}.

Before we formally define our reduction, we briefly describe the underlying
intuition.
The gadget in Figure~\ref{fig:complementary-reduction-example} is added for 
every complementary pair $(a,b)$.
It is constructed in such a way that every solution must either 
contain all of $\{a,a^{ab},a^{ab}_1,\dots,a^{ab}_{n+4}\}$ or none 
of them, and the same holds for 
$\{b,b^{ab},b^{ab}_1,\dots,b^{ab}_{n+4}\}$.
By making the vertex $\triangle^{ab}$ necessary, every solution must contain 
one of these two sets.
At the same time, the bound on the solution size makes sure that we cannot 
afford to take both sets for any complementary pair.

\begin{definition}
We define a function \tFNC{}, which assigns a \SSFN{} instance to each \SSFNC{} instance
$I = (G,k,V_\square,V_\triangle,C)$.
For this, we use
$n$ to denote $\size{V(G)}$
and first define a function
$\sFNC{I} : x \mapsto x + \size{C} \cdot (n+6)$.
For each $(a,b) \in C$, we introduce new vertices
$a^{ab}$, $b^{ab}$ and $\triangle^{ab}$ as well as,
for any $x \in \{a,b\}$, sets of new vertices
$Y^{ab}_{x\medcirc} = \{x^{ab}_1, \dots, x^{ab}_{n+1}\}$,
$Z^{ab}_{x\medcirc} = \{x^{ab}_{n+2}, x^{ab}_{n+3}, x^{ab}_{n+4}\}$,
$Y^{ab}_{x\square} = \{x^{ab\square}_1, \dots, x^{ab\square}_{n+1}\}$ and
$Z^{ab}_{x\square} = \{x^{ab\square}_{n+2}, x^{ab\square}_{n+3}, x^{ab\square}_{n+4}\}$.
We use the notation
$u \oplus v$
to denote the set of edges
$\{(u,v), (u,u^\square), (v,v^\square), (u,v^\square), (v,u^\square)\}$.
\input{complementary-reduction-figure}
Now we define
the \SSFN{} instance
$\tFNC(I) = (G',k',V_\square',V_\triangle')$, where
$k' = \sFNC{I}(k)$,
$V_\square' = V_\square \cup \bigcup_{(a,b) \in C} (Y^{ab}_{a\square} \cup Y^{ab}_{b\square} \cup Z^{ab}_{a\square} \cup Z^{ab}_{b\square})$,
$V_\triangle' = V_\triangle \cup \bigcup_{(a,b) \in C} \{\triangle^{ab}\}$
and $G'$ is the graph defined by
\[V(G') = V(G) \cup \bigcup_{(a,b) \in C} \big(\{\triangle^{ab}, a^{ab}, b^{ab}\} \cup Y^{ab}_{a\medcirc} \cup Y^{ab}_{b\medcirc} \cup Y^{ab}_{a\square} \cup Y^{ab}_{b\square} \cup Z^{ab}_{a\medcirc} \cup Z^{ab}_{b\medcirc} \cup Z^{ab}_{a\square} \cup Z^{ab}_{b\square}\big),\]
\begin{align*}
E(G') = E(G) \cup \bigcup_{(a,b) \in C}
\bigcup_{x \in \{a,b\}} \big(
&
\{(\triangle^{ab}, x^{ab})\}
\cup
(\{x\} \times Y^{ab}_{x\medcirc})
\cup
(\{x^{ab}\} \times Z^{ab}_{x\medcirc})
\\
{}\cup{}&
\bigcup_{1\leq i \leq n+3} x_i^{ab} \oplus x_{i+1}^{ab}
\big).
\end{align*}
We illustrate our construction in Figure~\ref{fig:complementary-reduction-example}.
\label{def:ssfnc-to-ssfn}
\end{definition}

\begin{lemma}
%
Let $I = (G,k,V_\square,V_\triangle,C)$ be a \SSFNC{} instance,
let $A$ be the set of
solutions of $I$
and let $B$ be the set of
solutions of $\tFNC(I)$.
There is a bijection $g: A \to B$
such that
$\size{g(S)} = \sFNC{I}(\size{S})$ holds for every $S \in A$.
\label{lem:ssfnc-to-ssfn-correct}
\end{lemma}

\begin{proof}
We use the same auxiliary notation as in Definition~\ref{def:ssfnc-to-ssfn}
and we define $g$ as
$S \mapsto S \cup \bigcup_{(a,b) \in C,\, x \in S \cap \{a,b\}} (\{\triangle^{ab}, x^{ab}\} \cup Y^{ab}_{x\medcirc} \cup Z^{ab}_{x\medcirc})$.
For every $S \in A$,
we thus obtain
$\size{g(S)} = \sFNC{I}(\size{S})$,
and we first show that indeed $g(S) \in B$.

Let $S \in A$ and let $S'$ denote $g(S)$.
Obviously $S'$ satisfies
$V_\square' \cap S' = \emptyset$ and
$V_\triangle' \subseteq S'$.
To see that $S'$ is secure in $G'$,
let $X'$ be an arbitrary subset of $S'$.
Since $S$ is secure in $G$ and $X' \cap V(G) \subseteq S$, there is a defense $\mu: N_G[X' \cap V(G)] \setminus S \to N_G[X' \cap V(G)] \cap S$.
We now construct a defense $\mu': N_{G'}[X'] \setminus S' \to N_{G'}[X'] \cap S'$.
For any attacker $v$ of $X'$ in $G'$, we distinguish three cases.
\begin{itemize}
\item If $v$ is some $x_i^{ab\square} \in Y^{ab}_{x\square} \cup Z^{ab}_{x\square}$ for some $(a,b) \in C$ and $x \in \{a,b\}$, we set $\mu'(v) = x_i^{ab}$.
This element is in $N_{G'}[X']$ since
$v$ is only adjacent to $x_i^{ab}$ or neighbors of it.
\item If $v$ is $a^{ab}$ or $b^{ab}$ for some $(a,b) \in C$, its only neighbor in $X'$ can be $\triangle^{ab}$ and we set $\mu'(v) = \triangle^{ab}$.
\item Otherwise $v$ is in $N_G[X' \cap V(G)] \setminus S$ (by our construction of $S'$).
Since the codomain of $\mu$ is a subset of the codomain of $\mu'$, we may set $\mu'(v) = \mu(v)$.
\end{itemize}
Since $\mu'$ is injective, each attack on $X'$ in $G'$ can be repelled by $S'$.
Hence $S'$ is secure in $G'$.

Clearly $g$ is injective.
It remains to show that $g$ is surjective.
Let $S'$ be a solution of $\tFNC(I)$.
First we make the following observations for each $(a,b) \in C$ and each $x \in \{a,b\}$:
\begin{itemize}
\item
If some $x^{ab}_i \in Y^{ab}_{x\medcirc}$ is in $S'$, then
$Y^{ab}_{x\medcirc} \cup Z^{ab}_{x\medcirc} \cup \{x\} \subseteq S'$
by Observation~\ref{obs:all-in-s}.
\item If some $x^{ab}_i \in Z^{ab}_{x\medcirc}$ is in $S'$, then $Y^{ab}_{x\medcirc} \cup Z^{ab}_{x\medcirc} \cup \{x^{ab}\} \subseteq S'$ for the same reason.
\item
If $x \in S'$, then $Y^{ab}_{x\medcirc} \cap S' \neq \emptyset$.
To see this, suppose $x \in S'$.
Let $D_x$ consist of those pairs $(c,d) \in C$ such that $x \in (c,d)$ and 
$Y^{cd}_{x\medcirc} \cap S' \neq \emptyset$,
and let $A_x$ consist of those pairs $(c,d) \in C$ such that $x \in (c,d)$ and 
$Y^{cd}_{x\medcirc} \cap S' = \emptyset$.
Now let $X' = \{x\} \cup \{x^{cd}_1,\dots,x^{cd}_n \mid (c,d) \in D_x\}$.
By the previous observations, $X' \subseteq S'$.
The defenders of $X'$ are the element $x$, the $\size{D_x} \cdot (n+1)$ 
elements of $\bigcup_{(c,d) \in D_x} Y^{cd}_{x\medcirc}$ and perhaps some 
elements of $N_G(x)$, which consists of at most $n-1$ vertices.
The attackers of $X'$ are the $\size{D_x} \cdot (n+1)$ elements of 
$\bigcup_{(c,d) \in D_x} Y^{cd}_{x\square}$, the $\size{A_x} \cdot (n+1)$ 
elements of $\bigcup_{(c,d) \in A_x} Y^{cd}_{x\medcirc}$ and perhaps some 
elements of $N_G(x)$.
Thus, if $A_x$ is nonempty, then the set $X'$ has more attackers than defenders 
in $G'$.
However, $S'$ is secure, so $A_x$ must be empty, which implies 
$Y^{ab}_{x\medcirc} \cap S' \neq \emptyset$.
\item If $x^{ab} \in S'$, then $Z^{ab}_{x\medcirc} \cap S' \neq \emptyset$
by Observation~\ref{obs:one-in-s}.
\end{itemize}
So for each $(a,b) \in C$ and $x \in \{a,b\}$,
$S'$ contains either all or none of
$\{x, x^{ab}\} \cup Y^{ab}_{x\medcirc} \cup Z^{ab}_{x\medcirc}$.

For every $(a,b) \in C$,
$S'$ contains $a^{ab}$ or $b^{ab}$,
since $\triangle^{ab} \in S'$,
whose neighbors are $a^{ab}$ and $b^{ab}$.
It follows that $\size{S'} > \size{C} \cdot (n+6)$ even if $S'$ contains only one of each $(a,b) \in C$.
If, for some $(a,b) \in C$, $S'$ contained both $a$ and $b$, we could derive a contradiction to $\size{S'} \leq \sFNC{I}(k) = k + \size{C} \cdot (n+6)$ because then $\size{S'} > (\size{C}+1) \cdot (n+6) > \sFNC{I}(k)$.
So $S'$ contains either $a$ or $b$ for any $(a,b) \in C$.

We construct $S = S' \cap V(G)$ and observe that
$S' = g(S)$,
$V_\triangle \subseteq S$, $V_\square \cap S = \emptyset$, and 
$\size{S \cap \{a,b\}} = 1$ for each $(a,b) \in C$.
It remains to show that $S$ is secure in $G$.
%
%
Let $X$ be an arbitrary subset of $S$.
We construct $X' = X \cup \bigcup_{(a,b) \in C, x \in X \cap \{a,b\}} Y^{ab}_{x\medcirc}$ and observe that
each $Y^{ab}_{x\medcirc}$ we put into $X'$ entails
$\size{Y^{ab}_{x\medcirc} \cup \{x_{n+2}^{ab}\}} = n+2$
additional defenders and
$\size{Y^{ab}_{x\square} \cup \{x_{n+2}^{ab\square}\}} = n+2$
additional attackers of
$X'$ in $G'$ compared to $X$ in $G$;
so $\size{N_{G'}[X'] \cap S'} - \size{N_G[X] \cap S} = \size{N_{G'}[X'] \setminus S'} - \size{N_G[X] \setminus S}$.
Clearly $X' \subseteq S'$, so $\size{N_{G'}[X'] \cap S'} \geq \size{N_{G'}[X'] \setminus S'}$ as $S'$ is secure in $G'$.
We conclude 
$  \size{N_G[X] \cap S} \geq \size{N_G[X] \setminus S}$.
Hence $S$ is secure in $G$.
\end{proof}

As $\tFNC$ is clearly computable in polynomial time, the following result follows:

\begin{corollary}
\SSFN{} is \Sptwo{}-hard.
\label{cor:ssfn-sptwo-hard}
\end{corollary}

The instances of \SSFNC{} are identical to the instances of the exact variant,
so $\tFNC$ is also applicable to the exact case.
In fact it turns out that this gives us also a reduction from \ESSFNC{} to \ESSFN{}.

\begin{corollary}
\ESSFN{} is \Sptwo{}-hard.
\label{cor:essfn-sptwo-hard}
\end{corollary}

\begin{proof}
Let $I$ and $I' = \tFNC(I)$ be our \ESSFNC{} and \ESSFN{} instances, respectively, and let
$k$ and $k'$ denote their respective solution sizes.
By Lemma~\ref{lem:ssfnc-to-ssfn-correct},
there is a bijection $g$ between the
solutions of $I$ and the solutions of $I'$
such that,
for every solution $S$ of $I$, $g(S)$ has $\sFNC{I}(k) = k'$ elements,
and for every solution $S'$ of $I'$,
$g^{-1}(S')$ has $k$ elements since $\sFNC{I}$ is invertible.
\end{proof}

\subsection{Hardness of Secure Set With Forbidden Vertices}

Now we present a transformation $\tFN$ that eliminates necessary vertices.
Our transformation not only operates on a problem instance, but also requires
an ordering $\preceq$ of the necessary vertices.
For now, we can consider this as an arbitrary ordering.
It will become more important in Section~\ref{sec:wone-treewidth}, where we
reuse this transformation for showing \Wone{}-hardness w.r.t.\ treewidth.

\input{necessary-reduction-example-figure}
Before formally defining the transformation $\tFN$, we refer to
Figure~\ref{fig:necessary-reduction-example-figure}, which shows the result
for a simple example graph with only two vertices $a$ and $b$, of which $b$ is
necessary.
The basic idea is that the vertex $w$ must be in every solution $S$ because
any vertex that is in $S$ also eventually forces $w$ to be in $S$.
Once $w \in S$, the construction to the right of $w$ makes sure that
$b \in S$.

\begin{definition}
We define a function \tFN{}, which assigns a \SSF{} instance to each pair $(I,{\preceq})$,
where
$I = (G,k,V_\square,V_\triangle)$ is a \SSFN{} instance and
${\preceq}$ is an ordering of the elements of $V_\triangle$.
For this,
let $V_\medcirc$ denote $V(G) \setminus (V_\square \cup V_\triangle)$.
We use
$n$ to denote $\size{V(G)}$,
and we first define a function
$\sFN{I} : x \mapsto xn + 3x + n - \size{V_\triangle} + \size{V_\medcirc} + 2$.
%
We use $W$ to denote the set of new vertices $\{w\} \cup \{w_v, w_v', w_v^\square, w_v'^\square \mid v \in V_\medcirc\}$.
The intention is for each $w_v^\square$ and $w_v'^\square$ to be forbidden, for $w$ and each $w_v'$ to be in every secure set, and for $w_v$ to be in a secure set iff $v$ is in it at the same time.
We write
$V^+$ to denote $V_\triangle \cup V_\medcirc \cup \{w\}$;
for each $v \in V^+$, we use $A_v$ to denote the set of new vertices
$\{v_1, \dots, v_{n+1}, v_1^\square, \dots, v_{n+1}^\square \}$,
and we use 
shorthand notation $A_v^\medcirc = \{v_1, \dots, v_{n+1}\}$ and $A_v^\square = \{v_1^\square, \dots, v_{n+1}^\square\}$.
The intention is for each $v_i^\square$ to be forbidden and for each $v_i$ to be in a secure set iff $v$ is in it at the same time.
We use the notation 
$u \oplus v$
to denote the set of edges
$\{(u,v), (u,u^\square), (v,v^\square), (u,v^\square), (v,u^\square)\}$.
%
If $V_\triangle = \emptyset$, let $P = \emptyset$; otherwise
let $P$ be the set consisting of all pairs $(u,v)$ such that $v$ is the direct
successor of $u$ according to ${\preceq}$,
as well as the pair $(u,w)$, where $u$ is the greatest element according to
${\preceq}$.
Now we define $\tFN(I,{\preceq}) = (G',k',V_\square')$, where
$V'_\square = V_\square \cup \{w_v^\square, w_v'^\square \mid v \in V_\medcirc\} \cup \bigcup_{v \in V^+} A_v^\square$,
$k' = \sFN{I}(k)$, and
$G'$ is the graph defined by
\begin{align*}
V(G') ={}& V(G) \cup W \cup \bigcup_{v \in V^+} A_v,\\
E(G') ={}& E(G) \cup \{ (v, v_i) \mid v \in V^+,\; 1 \leq i \leq n+1 \}\\
          {}\cup{}& \bigcup_{v \in V^+,\; 1 \leq i \leq n} v_i \oplus v_{i+1}
		   \cup \bigcup_{(u,v) \in P} u_{n+1} \oplus v_1\\
          {}\cup{}& \bigcup_{v \in V_\medcirc} v_{n+1} \oplus w_v
          \cup \{ (w,w_v), (w,w_v'), (w_v,w_v'), (w_v,w_v'^\square) \mid v \in V_\medcirc \}.
\end{align*}
We illustrate our construction in Figure~\ref{fig:necessary-reduction-gadget1} and~\ref{fig:necessary-reduction-gadget2}.
\label{def:ssfn-to-ssf}
\end{definition}

\begin{figure}
\centering
\begin{tikzpicture}
[scale=0.9,inner sep=2pt]
%
%
%
%
\node (a) at (2,2) {$a$};

\node (a1) at (0,1) {$a_1$};
\node (a2) at (1,1) {$a_2$};
\node (dots) at (2,1) {$\cdots$};
\node (an) at (3,1) {$a_n$};
\node (an1) at (4,1) {$a_{n+1}$};

\node (a1s) at (0,0) {$a_1^\square$};
\node (a2s) at (1,0) {$a_2^\square$};
\node (ans) at (3,0) {$a_n^\square$};
\node (an1s) at (4,0) {$a_{n+1}^\square$};

\node (wa) at (5,1) {$w_a$};
\node (was) at (5,0) {$w_a^\square$};
\node (waps) at (6,0) {$w_a'^\square$};
\node (wap) at (6,1) {$w_a'$};
\node (w) at (6,2) {$w$};

\draw (a) -- (a1);
\draw (a) -- (a2);
\draw (a) -- (an);
\draw (a) -- (an1);

\draw (w) -- (wa);
\draw (w) -- (wap);

\draw (a1) -- (a1s);
\draw (a2) -- (a2s);
\draw (an) -- (ans);
\draw (an1) -- (an1s);
\draw (wa) -- (was);

\draw (a1) -- (a2);
\draw (an) -- (an1) -- (wa) -- (wap);
\draw (a1) -- (a2s);
\draw (a2) -- (a1s);
\draw (an) -- (an1s);
\draw (an1) -- (ans);
\draw (an1) -- (was);
\draw (wa) -- (an1s);
\draw (waps) -- (wa);
\end{tikzpicture}
\caption{Illustration of the gadget that makes sure $w$ and $w_a'$ are in every secure set.
The vertex $a$ is a non-necessary, non-forbidden vertex from the \SSFN\ instance and may have other neighbors from this instance.
The vertex $w$ has two neighbors (as depicted here) for each non-necessary, non-forbidden vertex from the \SSFN\ instance, and additionally the neighbors depicted in Figure~\ref{fig:necessary-reduction-gadget2}.}
\label{fig:necessary-reduction-gadget1}
\end{figure}
\begin{figure}
\centering
\begin{tikzpicture}
[scale=0.9,inner sep=2pt]
\node (x) at (2,2) {$x$};

\node (x1) at (0,1) {$x_1$};
\node (x2) at (1,1) {$x_2$};
\node (xdots) at (2,1) {$\cdots$};
\node (xn) at (3,1) {$x_n$};
\node (xn1) at (4,1) {$x_{n+1}$};

\node (x1s) at (0,0) {$x_1^\square$};
\node (x2s) at (1,0) {$x_2^\square$};
\node (xns) at (3,0) {$x_n^\square$};
\node (xn1s) at (4,0) {$x_{n+1}^\square$};

\node (y) at (7,2) {$y$};

\node (y1) at (5,1) {$y_1$};
\node (y2) at (6,1) {$y_2$};
\node (ydots) at (7,1) {$\cdots$};
\node (yn) at (8,1) {$y_n$};
\node (yn1) at (9,1) {$y_{n+1}$};

\node (y1s) at (5,0) {$y_1^\square$};
\node (y2s) at (6,0) {$y_2^\square$};
\node (yns) at (8,0) {$y_n^\square$};
\node (yn1s) at (9,0) {$y_{n+1}^\square$};

\node (w) at (12,2) {$w$};

\node (w1) at (10,1) {$w_1$};
\node (w2) at (11,1) {$w_2$};
\node (wdots) at (12,1) {$\cdots$};
\node (wn) at (13,1) {$w_n$};
\node (wn1) at (14,1) {$w_{n+1}$};

\node (w1s) at (10,0) {$w_1^\square$};
\node (w2s) at (11,0) {$w_2^\square$};
\node (wns) at (13,0) {$w_n^\square$};
\node (wn1s) at (14,0) {$w_{n+1}^\square$};

\draw (x1) -- (x2) -- (x1s);
\draw (x1) -- (x2s);
\draw (xn) -- (xn1) -- (y1) -- (y2) -- (y1s) -- (xn1) -- (xns);
\draw (xn) -- (xn1s) -- (y1) -- (y2s);
\draw (yn) -- (yn1) -- (w1) -- (w2) -- (w1s) -- (yn1) -- (yns);
\draw (yn) -- (yn1s) -- (w1) -- (w2s);
\draw (wn) -- (wn1) -- (wns);
\draw (wn) -- (wn1s);

\foreach \x in {x,y,w} {
	\foreach \y in {1,2,n,n1} {
		\draw (\x) -- (\x\y) -- (\x\y s);
	}
}
\end{tikzpicture}
\caption{Illustration of the gadget that makes sure every secure set contains all necessary vertices as it must contain $w$.
Here we assume there are the two necessary vertices $x$ and $y$, and we use the ordering $x \preceq y$.}
\label{fig:necessary-reduction-gadget2}
\end{figure}

We now prove that $\tFN$ yields a correct reduction for any ordering $\preceq$.

\begin{lemma}
Let $I = (G,k,V_\square,V_\triangle)$ be a \SSFN{} instance,
let ${\preceq}$ be an ordering of $V_\triangle$,
let $A$ be the set of solutions of $I$ and
let $B$ be the set of solutions of $\tFN(I,{\preceq})$.
There is a bijection $g: A \to B$ such that
$\size{g(S)} = \sFN{I}(\size{S})$
holds for every $S \in A$.
\label{lem:ssfn-to-ssf-correct}
\end{lemma}

\begin{proof}
We use the same auxiliary notation as in Definition~\ref{def:ssfn-to-ssf}
and we define $g$ as
$S \mapsto S
\cup \bigcup_{v \in S} A_v^\medcirc
\cup A_w^\medcirc
\cup \{w\}
\cup \{w_v' \mid v \in V_\medcirc\}
\cup \{w_v \mid v \in S \cap V_\medcirc\}
$.
For every $S \in A$,
we thus obtain
$\size{g(S)}
= \size{S} + \size{S} (n+1) + (n+1) + 1 + \size{V_\medcirc} + (\size{S}-\size{V_\triangle})
= \sFN{I}(\size{S})$,
and we first show that indeed $g(S) \in B$.

Let $S \in A$ and let
$S'$ denote $g(S)$.
Obviously $S'$ satisfies
$V_\square' \cap S' = \emptyset$.
To see that $S'$ is secure in $G'$,
let $X'$ be an arbitrary subset of $S'$.
Since $S$ is secure in $G$ and $X' \cap V(G) \subseteq S$, there is a defense $\mu: N_G[X' \cap V(G)] \setminus S \to N_G[X' \cap V(G)] \cap S$.
We now construct a defense $\mu': N_{G'}[X'] \setminus S' \to N_{G'}[X'] \cap S'$.
For any attacker $a$ of $X'$ in $G'$, we distinguish the following cases:
\begin{itemize}
\item If $a$ is some $v_i^\square \in A_v^\square$ for some $v \in V^+$,
then $a$ can only attack either $v_i$ or a neighbor of $v_i$, all of which are in $S'$,
and we set $\mu'(a) = v_i$.
\item Similarly, if $a$ is $w_v^\square$ for some $v \in V_\medcirc$,
then we set $\mu'(a) = w_v$.
\item If $a$ is $w_v'^\square$ for some $v \in V_\medcirc$,
then $a$ attacks $w_v$ and we set $\mu'(a) = w_v'$.
\item If $a$ is $w_v$ for some $v \in V_\medcirc$,
then it attacks $w$ or $w_v'$,
which is not used for repelling any attack because $w_v'^\square$ cannot attack $X'$,
so we set $\mu'(a) = w_v'$.
\item Otherwise $a$ is in $N_G[X' \cap V(G)] \setminus S$ (by our construction of $S'$).
Since the codomain of $\mu$ is a subset of the codomain of $\mu'$, we may set $\mu'(a) = \mu(a)$.
\end{itemize}
Since $\mu'$ is injective, each attack on $X'$ in $G'$ can be repelled by $S'$.
Hence $S'$ is secure in $G'$.

Clearly $g$ is injective.
It remains to show that $g$ is surjective.
Let $S'$ be a solution of $\tFN(I,{\preceq})$.
We first show that $V_\triangle \cup \{w\} \subseteq S'$:
\begin{itemize}
\item If $S'$ contains some $v \in V_\triangle \cup V_\medcirc$, then
$S'$ contains an element of $A_v^\medcirc$
by Observation~\ref{obs:one-in-s}.
\item If $S'$ contains an element of $A_v^\medcirc$ for some $v \in V^+$, then
$\{v\} \cup A_v^\medcirc \subseteq S'$
by Observation~\ref{obs:all-in-s}.
\item If $v_{n+1} \in S'$ for some $v \in V_\medcirc$, then
$w_v \in S'$ for the same reason.
\item Furthermore, if $S'$ contains an element of $A_v^\medcirc$ for some $v \in V_\triangle \cup \{w\}$, then
also $A_{u}^\medcirc \subseteq S'$ for every $u \in V_\triangle \cup \{w\}$
for the same reason.
\item If $w_v \in S'$ for some $v \in V_\medcirc$, then
$\{w,w_v',v_{n+1}\} \subseteq S'$
by Observation~\ref{obs:all-in-s}.
\item If $w_v' \in S'$ for some $v \in V_\medcirc$, then
$w \in S'$ because at least $w_v$ or $w$ must be in $S'$
and the former implies $w \in S'$ as we have seen.
\item The previous observations show that any vertex being in $S'$ implies $w \in S'$.
Since $S'$ is nonempty,
it follows that $w \in S'$.
We now show that $S'$ contains an element of $A_w^\medcirc$.
Suppose the contrary, let
$U = S' \cap \{w_v \mid v \in V_\medcirc\}$, let
$U' = S' \cap \{w_v' \mid v \in V_\medcirc\}$
and consider
$X' = \{w\} \cup U$.
The defenders of $X'$
consist of exactly
$1 + \size{U'} + 2 \size{U}$ elements,
whereas there are exactly
$(n+1) + (\size{V_\medcirc}- \size{U'}) + (\size{V_\medcirc} - \size{U}) + 3 \size{U}$
attackers.
With $\size{V_\medcirc} \geq \size{U'} \geq \size{U}$ and $n > 0$ in mind,
we arrive at the contradiction
$\size{N_{G'}[X'] \cap S'} < \size{N_{G'}[X'] \setminus S'}$.
\item
The previous observations show that
for every $v \in V_\triangle \cup \{w\}$
it holds that
$\{v\} \cup A_v^\medcirc \subseteq S'$.
Finally, we show that $\{w_v' \mid v \in V_\medcirc\} \subseteq S'$.
Suppose, for the sake of contradiction, that there is some
$u \in V_\medcirc$ such that
$w_u' \notin S'$.
We have seen that the latter can only be the case if $u \notin S'$.
Observe that
$\{w\} \cup \{w_i \mid 2 \leq i \leq n+1\} \cup \{w_v \mid v \in V_\medcirc \cap S'\}$
is a subset of $S'$ that is attacked by
$\{w_u'\} \cup
A_w^\square \cup
\{v_{n+1}^\square, w_v^\square, w_v'^\square \mid v \in V_\medcirc \cap S'\} \cup
\{w_v \mid v \in V_\medcirc \setminus S'\}$,
but the defenders are a proper subset of
$\{w\} \cup
A_w^\medcirc \cup
\{v_{n+1}, w_v, w_v' \mid v \in V_\medcirc \cap S'\} \cup
\{w_v' \mid v \in V_\medcirc \setminus S'\}$.
This contradicts $S'$ being secure in $G'$.
\end{itemize}
%
Let $S = S' \cap V(G)$.
By the previous observations, it is easy to see that $S' = g(S)$.
It remains to show that $S$ is secure in $G$.
Let $X$ be an arbitrary subset of $S$.
We construct $X' = X \cup \bigcup_{v \in X} A^\medcirc_v$ and observe that
the number of additional defenders of $X'$ in $G'$ compared to $X$ in $G$ is equal to
the number of additional attackers;
formally $\size{N_{G'}[X'] \cap S'} - \size{N_G[X] \cap S} = \size{N_{G'}[X'] \setminus S'} - \size{N_G[X] \setminus S}$.
Clearly $X' \subseteq S'$, so
$\size{N_{G'}[X'] \cap S'} \geq \size{N_{G'}[X'] \setminus S'}$
as $S'$ is secure in $G'$.
Consequently
$\size{N_G[X] \cap S} \geq \size{N_G[X] \setminus S}$.
Hence $S$ is secure in $G$.
%
\end{proof}

Given an ordering ${\preceq}$, clearly $\tFN(I,\preceq)$ is computable in polynomial time.
We can thus easily obtain a reduction from \SSFN{} to \SSF{}
by first computing an arbitrary ordering ${\preceq}$ of the necessary vertices
in polynomial time.
This also gives us a hardness result for the exact case, analogous to
Corollary~\ref{cor:essfn-sptwo-hard}.

\begin{corollary}
\SSF{} and \ESSF{} are \Sptwo{}-hard.
\label{cor:ssf-sptwo-hard}
\end{corollary}

\subsection{Hardness of Secure Set}

We now introduce a transformation $\tF$ that eliminates forbidden vertices.
The basic idea is that we ensure that a forbidden vertex $f$ is never part of a 
solution by adding so many neighbors to $f$ that we could only defend $f$ by 
exceeding the bound on the solution size.

\begin{definition}
We define a function \tF{}, which assigns a \probSS{} instance to each \SSF{} instance
$I = (G,k,V_\square)$.
\label{def:ssf-to-ss}
For each $f \in V_\square$, we
introduce new vertices $f', f_1, \dots, f_{2k}$.
Now we define $\tF(I) = (G',k)$, where
$G'$ is the graph defined by
\begin{align*}
V(G') ={}& V(G) \cup \{f', f_1, \dots, f_{2k} \mid f \in V_\square\},\\
E(G') ={}& E(G) \cup \{ (f,f_i),\; (f',f_i) \mid f \in V_\square,\; 1 \leq i \leq 2k\}.
\end{align*}
\end{definition}


\begin{lemma}
Every \SSF{} instance $I$ has the same solutions as the \probSS{} instance $\tF(I)$.
\label{lem:ssf-to-ss-correct}
\end{lemma}

\begin{proof}
Let $I = (G,k,V_\square)$ and $\tF(I) = (G',k)$.
Each secure set $S$ in $G$ is also secure in $G'$ because the subgraph of $G$ induced by $N_G[S]$ is equal to the subgraph of $G'$ induced by $N_{G'}[S]$.
Now let $S'$ be a solution of $\tF(I)$.
For every $f \in V_\square$,
neither $f$ nor $f'$ are in $S'$
because each of these vertices has at least $2k$ neighbors,
and $S'$ cannot contain any $f_i$ because $N_{G'}(f_i) = \{f,f'\}$.
Hence $S'$ is also secure in $G$ as the subgraphs induced by the respective neighborhoods are again equal.
\end{proof}

This immediately yields the following result.

\begin{corollary}
\probSS{} and \ESS{} are \Sptwo{}-hard.
\label{cor:ss-sptwo-hard}
\end{corollary}

\section{Complexity of Secure Set Parameterized by Treewidth}
\label{sec:parameterized-complexity-treewidth}

In this section we study the parameterized complexity of the \probSS{}
problem when treewidth is the parameter.

We first show that all variants of \probSS{} considered in this paper are
\Wone{}-hard for this parameter by reusing some reductions from
Section~\ref{sec:classical-complexity} and proving that they preserve bounded
treewidth.
Under the widely held assumption that $\FPT \neq \Wone$, this rules out
fixed-parameter tractable algorithms for these problems.

Second, we show that the
\CONP{}-complete \SSV{} problem is solvable in linear time on instances whose
treewidth is bounded by a constant.
We do this by providing a fixed-parameter linear algorithm that performs
dynamic programming on a tree decomposition of the input graph.
Although bounded treewidth most likely does not lead to fixed-parameter
tractability of the problem of \emph{finding} secure sets, this proves that it
does for the problem of verifying whether a given set is secure.

Third, we show that all the variants of the \probSS{} problem considered in
this paper are solvable in polynomial time on instances whose treewidth is
bounded by a constant. We again do this by providing a polynomial-time dynamic
programming algorithm, but this time the degree of the polynomial depends on
the treewidth.

\subsection{Hardness of Secure Set Parameterized by Treewidth}
\label{sec:wone-treewidth}

In this subsection, we prove the following theorem:
\begin{theorem}
\label{thm:ss-variants-wone}
The following problems are all $\Wone$-hard when parameterized by treewidth:
\probSS{},
\ESS{},
\SSF{},
\ESSF{},
\SSFN{},
\ESSFN{},
\SSFNC{}, and
\ESSFNC{}.
\end{theorem}
To prove this,
we reduce from the following problem~\cite{DBLP:journals/dam/AsahiroMO11},
which is known to be \Wone-hard~\cite{DBLP:journals/corr/abs-1107-1177} parameterized by the treewidth of the graph:
\decisionProblem{\MMO}{%
A graph $G$,
an edge weighting $w: E(G) \to \mathbb{N}^+$ given in unary
and a positive integer $r$}{%
Is there an orientation of the edges of $G$ such that, for each $v \in V(G)$, the sum of the weights of outgoing edges from $v$ is at most $r$?%
}

\begin{lemma}
\SSFNC{} and \ESSFNC{}, both parameterized by the treewidth of the primal graph, are \Wone-hard.
\end{lemma}

\begin{proof}
Let an instance of \MMO\ be given by a graph $G$, an edge weighting $w: E(G) \to \mathbb{N}^+$ in unary and a positive integer $r$.
From this we construct an instance of both \SSFNC{} and \ESSFNC{}.
An example is given in Figure~\ref{fig:minmaxoutdegree-reduction-example}.
For each $v \in V(G)$, we define the set of new vertices $H_v = \{h^v_1, \dots, h^v_{r-1}\}$, and
for each $(u,v) \in E(G)$, we define the sets of new vertices
$V_{uv} = \{u^v_1, \dots, u^v_{w(u,v)}\}$ and
$V_{vu} = \{v^u_1, \dots, v^u_{w(u,v)}\}$.
We now define the graph $G'$ with
\begin{align*}
V(G') ={}& V(G) \cup \bigcup_{v \in V(G)} H_v \cup \bigcup_{(u,v) \in E(G)} (V_{uv} \cup V_{vu}),\\
E(G') ={}& \{(v,h) \mid v \in V(G),\; h \in H_v\}\\
   {}\cup{}&\{(u,x) \mid (u,v) \in E(G),\; x \in V_{uv}\} \cup \{(x,v) \mid (u,v) \in E(G),\; x \in V_{vu}\}\\
   {}\cup{}&\{(u^v_i,v^u_i) \mid (u,v) \in E(G),\; 1 \leq i \leq w(u,v)\}.
\end{align*}
We also define the set of complementary vertex pairs
$C = \{(u^v_i,v^u_i) \mid (u,v) \in E(G),\; 1 \leq i \leq w(u,v)\} \cup
\{(v^u_i,u^v_{i+1}) \mid (u,v) \in E(G),\; 1 \leq i < w(u,v)\}$.
Finally, we define the set of necessary vertices
$V_\triangle = V(G) \cup \bigcup_{v \in V(G)} H_v$
and
$k = \size{V_\triangle} + \sum_{(u,v) \in E(G)} w(u,v)$.
We use $I$ to denote $(G',k,C,V_\triangle,\emptyset)$,
which is an instance of \SSFNC\ and also of \ESSFNC.
Obviously $I$ is a positive instance of \SSFNC\ iff
it is a positive instance of \ESSFNC\ because
the necessary and complementary vertices make sure that
every solution of the \SSFNC\ instance $I$ has exactly $k$ elements.
Hence we only consider \SSFNC.

\begin{figure}
\centering
\begin{tikzpicture}[node distance=1mm and 8mm,bend angle=15]
\node (a) [necessary,label=$a$] {};
\node (ab2) [right=of a] {$a^b_2$};
\node (ab1) [above=of ab2] {$a^b_1$};
\node (ab3) [below=of ab2] {$a^b_3$};
\node (ba1) [right=of ab1] {$b^a_1$};
\node (ba2) [right=of ab2] {$b^a_2$};
\node (ba3) [right=of ab3] {$b^a_3$};
\node (b) [necessary,right=of ba2,label=$b$] {};

\node (ha1) [necessary,above left=of a,label=$h^a_1$] {};
\node (ha2) [necessary,below left=of a,label=below:$h^a_2$] {};
\node (hb1) [necessary,above right=of b,label=$h^b_1$] {};
\node (hb2) [necessary,below right=of b,label=below:$h^b_2$] {};

\draw (a) -- (ha1);
\draw (a) -- (ha2);
\draw (a) -- (ab1);
\draw (a) -- (ab2);
\draw (a) -- (ab3);

\draw (b) -- (hb1);
\draw (b) -- (hb2);
\draw (b) -- (ba1);
\draw (b) -- (ba2);
\draw (b) -- (ba3);

\draw [dashed] (ab1) to (ba1);
\draw [dashed] (ab2) to (ba2);
\draw [dashed] (ab3) to (ba3);

\draw [bend left] (ab1) to (ba1);
\draw [bend left] (ab2) to (ba2);
\draw [bend left] (ab3) to (ba3);

\draw [dashed] (ba1) -- (ab2);
\draw [dashed] (ba2) -- (ab3);
\end{tikzpicture}
\caption{%
Result of our transformation on a sample \MMO\ instance with
$r=3$ and two vertices $a,b$ that are connected by an edge of weight~$3$.
Complementary vertex pairs are shown via dashed lines.
Necessary vertices have a $\triangle$ symbol next to their name.%
}
\label{fig:minmaxoutdegree-reduction-example}
\end{figure}

The intention is that for each orientation of $G$ we have a solution
candidate $S$ in $I$ such that
an edge orientation from $u$ to $v$ entails
$V_{vu} \subseteq S$ and
$V_{uv} \cap S = \emptyset$,
and the other orientation entails $V_{uv} \subseteq S$ and $V_{vu} \cap S = \emptyset$.
For each outgoing edge of $v$ in the orientation of $G$,
there are as many attackers of $v$ in $I$ as the weight of that edge.
Together with $H_v$, $v$ can repel up to $r$ such attacks.
The other neighbors of $v$ that are in $S$ cannot help $v$ since they are in
turn attacked by their neighbors.

Clearly $I$ can be computed in polynomial time.
We now show that the treewidth of the primal graph of $I$ depends only on the treewidth of $G$.
We do so by modifying an optimal tree decomposition $\T$ of $G$ as follows:
\begin{enumerate}
\item For each $(u,v) \in E(G)$, we take an arbitrary node whose bag $B$ contains
both $u$ and $v$
and add to its children
a chain of nodes $N_1, \dots, N_{w(u,v)-1}$ such that the
bag of $N_i$ is $B \cup \{u^v_i, u^v_{i+1}, v^u_i, v^u_{i+1}\}$.
\item For each $v \in V(G)$, we take an arbitrary node whose bag $B$ contains $v$
and add to its children
a chain of nodes
$N_1, \dots, N_{r-1}$
such that the bag of $N_i$ is
$B \cup \{h^v_i\}$.
\end{enumerate}
It is easy to verify that the result is a valid tree decomposition of the
primal graph of $I$ and its width is at most the treewidth of $G$ plus four.

We claim that $(G,w,r)$ is a positive instance of \MMO{} iff
$I$ is a positive instance of \SSFNC{}.

\medskip
\noindent
\emph{``Only if'' direction.}
Let $D$ be the directed graph given by an orientation of the edges of $G$ such that for each vertex the sum of weights of outgoing edges is at most $r$.
The set
$S = V_\triangle \cup \{v^u_1, \dots, v^u_{w(u,v)} \mid (u,v) \in E(D)\}$
is secure in $G$:
Let $X$ be an arbitrary subset of $S$.
Every attacker must be some element $u^v_i$.
If $v^u_i \in X$, then we can use $v^u_i$ to repel the attack from $u^v_i$.
Otherwise $u \in X$, so we can use either $u$ or one of the $r-1$ elements of $H_u$ to repel the attack from $u^v_i$.
These are sufficiently many defenders:
For every vertex $v \in V(G) \cap X$, at most $r$ neighbors attack $v$ as otherwise the sum of weights of outgoing edges of $v$ in $D$ would be greater than $r$.
Finally, it is easy to verify that
$\size{S} = k$,
$V_\triangle \subseteq S$, and exactly one element of each pair of complementary vertices is in $S$.

\medskip
\noindent
\emph{``If'' direction.}
Let $S$ be a solution of $I$.
For every $(u,v) \in E(G)$, either $V_{uv} \subseteq S$ or $V_{vu} \subseteq S$ due to the complementary vertex pairs.
We define a directed graph $D$ by $V(D) = V(G)$ and
$E(D) = \{(u,v) \mid V_{vu} \subseteq S\} \cup \{(v,u) \mid V_{uv} \subseteq S\}$.
Suppose there is a vertex $v$ in $D$ whose sum of weights of outgoing edges is greater than $r$.
We construct $X = \{v\} \cup \bigcup_{(u,v) \in E(D)} V_{vu}$, which is a subset of $S$.
Now $v$ has more than $r$ attacking neighbors, but all defenders except $v$ and $H_v$ must already defend themselves against their attacking neighbor.
This contradicts $S$ being secure.
\end{proof}

Now we reduce from \SSFNC{} to \SSFN{} to show \Wone{}-hardness of the latter
problem.
We reuse the function $\tFNC$ from Definition~\ref{def:ssfnc-to-ssfn} and show
that this gives us a reduction that preserves bounded treewidth.

\begin{lemma}
\SSFN{}, parameterized by the treewidth of the graph, is \Wone{}-hard.
\label{lem:ssfn-wone-hard}
\end{lemma}

\begin{proof}
Let $I$ be a \SSFNC{} instance whose primal graph we denote by $G$.
We obtain an equivalent \SSFN{} instance $\tFNC(I)$, whose graph we denote by $G'$.
This reduction is correct, as shown in Lemma~\ref{lem:ssfnc-to-ssfn-correct}.
It remains to show that the treewidth of $G'$ is bounded by a function of the treewidth of $G$.
Let $\T$ be an optimal nice tree decomposition of $G$.
We build a tree decomposition $\T'$ of $G'$ by modifying a copy of $\T$ in the 
following way:
For every pair $(a,b)$ of complementary vertices, we pick an arbitrary node $t$ 
in $\T$ whose bag $B$ contains both $a$ and $b$, and we add a chain of nodes 
$N_1, \dots, N_{2n+3}$ between $t$ and its parent such that,
for $1 \leq i \leq n+1$, the bag of $N_i$ is $B \cup \{a_i^{ab}, a_i^{ab\square}, a_{i+1}^{ab}, a_{i+1}^{ab\square}\}$,
the bag of $N_{n+2}$ is
$B \cup \{a^{ab}, b^{ab}, \triangle^{ab}\} \cup
Z_{a\medcirc}^{ab} \cup Z_{a\square}^{ab} \cup Z_{b\medcirc}^{ab} \cup Z_{b\square}^{ab}$,
and
the bag of $N_{n+2+i}$ is $B \cup \{b_{n+3-i}^{ab}, b_{n+3-1}^{ab\square}, b_{n+2-i}^{ab}, b_{n+2-i}^{ab\square}\}$.
It is easy to verify that $\T'$ is a valid tree decomposition of $G'$.
Furthermore, the width of $\T'$ is at most the width of $\T$ plus~15.
\end{proof}

Just like before, we get an analogous result for the exact variant.
It can be proved in the same way as Corollary~\ref{cor:essfn-sptwo-hard}.

\begin{corollary}
\ESSFN{}, parameterized by the treewidth of the graph, is \Wone{}-hard.
\label{cor:essfn-wone-hard}
\end{corollary}

We next show \Wone{}-hardness of \SSF{} by reducing from \SSFN{} using the
function $\tFN$ from Definition~\ref{def:ssfn-to-ssf}.
This function maps a \SSFN{} instance, together with an ordering $\preceq$ of
the necessary vertices, to an equivalent \SSF{} instance.
We show that by choosing $\preceq$ appropriately, this gives us a reduction
that preserves bounded treewidth.

\begin{lemma}
\SSF{}, parameterized by the treewidth of the graph, is \Wone{}-hard.
\label{lem:ssf-wone-hard}
\end{lemma}

\begin{proof}
Let $I = (G,k,V_\square,V_\triangle)$ be a \SSFN{} instance
and let $\T$ be an optimal nice tree decomposition of $G$.
We can compute such a tree decomposition in
FPT time~\cite{DBLP:journals/siamcomp/Bodlaender96}.
Let 
${\preceq}$
be 
the
ordering of the elements of $V_\triangle$
that is obtained in linear time by doing a post-order traversal of $\T$ and 
sequentially recording the elements that occur for the last time in the current 
bag.
We obtain the \SSF{} instance $\tFN(I,\preceq)$, whose graph we denote by $G'$.
This reduction is correct, as shown in Lemma~\ref{lem:ssfn-to-ssf-correct},
and computable in FPT time.
It remains to show that the treewidth of $G'$ is bounded by a function of the treewidth of $G$.
To this end, we use $\T$ to build a tree decomposition $\T'$ of $G'$.
We initially set $\T' := \T$ and modify it by the following steps:
\begin{enumerate}
\item We insert $w$ into every bag.
\item For every $(u,v) \in P$,
we add $v$, $v_1$ and
$v_1^\square$ into the bag of every node between (and including) $t_u^{\T'}$ and
$t_v^{\T'}$.
Note that
the bag of $t_u^{\T'}$ contains both $u$ and $v$.
After this step, we have increased the bag size of each node by at most five.
\item For each $v \in V^+$, we use $B_v$ to denote the bag of
$t_v^{\T'}$ and replace $t_v^{\T'}$ by a chain of nodes
$N_1, \dots, N_n$, where $N_n$ is the topmost node and the bag of
$N_i$ is $B_v \cup \{v_i, v_i^\square, v_{i+1}, v_{i+1}^\square\}$.
After this step,
note that, for each $(u,v) \in P$,
the bag of the new node $t_u^{\T'}$ contains $u_{n+1}$, $u_{n+1}^\square$, $v_1$ and $v_1^\square$.
\item For each $v \in V_\medcirc$, we
add $w_v$, $w_v^\square$, $w_v'$ and $w_v'^\square$ to the bag of $t_v^{\T'}$,
which already contains $w$, $v_{n+1}$, $v_{n+1}^\square$.
\end{enumerate}
It is easy to verify that $\T'$ is a valid tree decomposition of $G'$.
Furthermore, the width of $\T'$ is at most the width of $\T$ plus twelve.
\end{proof}

We again get an analogous result for the exact variant.

\begin{corollary}
\ESSF{}, parameterized by the treewidth of the graph, is \Wone{}-hard.
\label{cor:essf-wone-hard}
\end{corollary}

Finally, we show \Wone{}-hardness of \probSS{} by reducing from \SSF{} while
preserving bounded treewidth.

\begin{lemma}
\probSS{}, parameterized by the treewidth of the graph, is \Wone{}-hard.
\label{lem:ss-wone-hard}
\end{lemma}

\begin{proof}
Let $I = (G,k,V_\square)$ be a \SSF{} instance,
let $G'$ denote the graph of $\tF(I)$
and let $\T$ be an optimal nice tree decomposition of $G$.
We build a tree decomposition $\T'$ of $G'$ by modifying a copy of $\T$ in the following way:
For every $f \in V_\square$, we pick an arbitrary node $t$ in $\T$ whose bag $B$ contains $f$,
and we add a chain of nodes $N_1, \dots, N_{2k}$ between $t$ and its parent such that,
for $1 \leq i \leq 2k$, the bag of $N_i$ is
$B \cup \{f', f_i\}$.
It is easy to verify that $\T'$ is a valid tree decomposition of $G'$.
Furthermore, the width of $\T'$ is at most the width of $\T$ plus two.
\end{proof}

We again get an analogous result for the exact variant.

\begin{corollary}
\ESS{}, parameterized by the treewidth of the input graph, is \Wone{}-hard.
\label{cor:ess-wone-hard}
\end{corollary}

\subsection{A Fixed-Parameter Tractable Algorithm for Secure Set Verification}
\label{sec:fpt-verification}

While we have seen in Section~\ref{sec:wone-treewidth} that \probSS{}
parameterized by treewidth is most likely not FPT, we now present a positive
result:
The \CONP{}-complete~\cite{Ho2011} \SSV{} problem, which consists of checking
whether a given set $\hatS$ is secure in a graph $G$,
is FPT parameterized by the treewidth of $G$.
We show this by giving a fixed-parameter linear algorithm that follows the
principle of dynamic programming on a tree decomposition $\T$ of $G$.
The core idea is the following:
For each node $t$ of $\T$ and each $X \subseteq \hatS \cap \chi(t)$, we store
an integer $\cvalhatStX$, which indicates that $X$ can be extended to a set
$\hatX \subseteq \hatS$ using ``forgotten'' vertices from further down in $\T$
in such a way that the difference between defenders and attackers of $\hatX$ is
$\cvalhatStX$ and $\hatX$ is the ``worst'' subset of $\hatS$ that can be
obtained in this way.
To compute these values, we traverse $\T$ from the bottom up and use recurrence
relations to compute the values for the current node $t$ of $\T$ based on the
values we have computed for the children of $t$.
If we then look at the values we have computed at the root of $\T$, we can
decide if there is a subset of $\hatS$ that is ``bad enough'' to witness that
$\hatS$ is not secure.

Dynamic programming algorithms like this are quite common for showing \FPT{}
membership w.r.t.\ treewidth and some examples can be found
in~\cite{niedermeier2006invitation}.
Proving their correctness is a usually rather tedious structural induction
argument along the tree decomposition:
At every node $t$ of $\T$, we have to prove that the recurrence relations
indeed characterize the value they are supposed to represent.
Examples of such proofs can be found in~\cite{Cygan15}.

We now formally define the values that we will compute at each tree
decomposition node.
Let $G$ be a graph with a nice tree decomposition $\T$ and let
$\hatS \subseteq V(G)$ be the candidate for which we want to check if it is secure.
For each node $t$ of $\T$ and each set of vertices $A$, we define
$A_t = \{a \in A \mid a \in \chi(t'),\, t' \text{ is a descendant of } t\}$.
For any $\hatX \subseteq \hatS$, we call $\size{N_G[\hatX]_t \cap \hatS} - \size{N_G[\hatX]_t \setminus \hatS}$ the
\emph{score} of $\hatX$ w.r.t.\ $\hatS$ at $t$ (or just the score of $\hatX$ if $\hatS$ and $t$
are clear from the context) and denote it by $\scorehatSthatX$.
Furthermore, we call
$\size{N_G[\hatX] \cap \chi(t) \cap \hatS} - \size{(N_G[\hatX] \cap \chi(t)) \setminus \hatS}$
the \emph{local score} of $\hatX$ w.r.t.\ $\hatS$ at $t$
and denote it by $\lscorehatSt(\hatX)$.
Finally,
for each $X \subseteq \hatS \cap \chi(t)$,
we define the value
\[\cvalhatStX = \min_{\hatX \subseteq \hatSt,\; \hatX \cap \chi(t) = X}\{\scorehatSthatX\}.\]
When $r$ is the root of $\T$, both $\hatS_r = \hatS$ and $\chi(r) = \emptyset$ hold, so
$\hatS$ is secure if and only if $\cval{\hatS}{r}(\emptyset)$ is nonnegative.

We now describe how to compute all such values in a bottom-up manner by
distinguishing the node type of $t$, and we prove the correctness of our
computation by structural induction along the way.
In this correctness proof, we use additional terminology:
We say that a set $\hatX$ is an \emph{extension} of $X$ w.r.t.\ $\hatS$ at $t$ if it is one of those sets considered in the definition of $\cvalhatStX$ that has minimum score; formally
$\hatX \subseteq \hatSt$, $\hatX \cap \chi(t) = X$ and $\score{\hatS}{t}(\hatX) = \cvalhatStX$.
We may omit $\hatS$ or $t$ if they are clear from the context.

\begin{description}
\item[Leaf node.]
If $t$ is a leaf node, then its bag is empty and obviously $\cvalhatSt(\emptyset) = 0$ holds.

\item[Introduce node.]
Let $t$ be an introduce node with child $t'$,
let $v$ be the unique element of $\chi(t) \setminus \chi(t')$,
let $X \subseteq \hatS \cap \chi(t)$ and
let $X' = X \setminus \{v\}$.
We prove that the following equation holds:
\[\cvalhatStX = \begin{cases}
    \cvalhatStpXp + 1 &\text{if } v \in N_G[X] \cap \hatS\\
    \cvalhatStpXp - 1 &\text{if } v \in N_G[X] \setminus \hatS\\
    \cvalhatStpXp &\text{otherwise}
\end{cases}\]

First consider the case where $v \in N_G[X] \cap \hatS$.
Let $\hatX$ be an extension of $X$ at $t$,
so $\scorehatSthatX = \cvalhatStX$.
From
$v \notin N_G[\hatX \setminus \{v\}]_{t'}$ and
$v \in N_G[\hatX]_t \cap \hatS$
we infer
$\scorehatSthatX = \scorehatStp(\hatX \setminus \{v\}) + 1$.
Moreover, the set $\hatX \setminus \{v\}$ is one of the candidates considered for an extension of $X'$ in the definition of $\cvalhatStp$, so we obtain
$\cvalhatStpXp \leq \score{\hatS}{t'}(\hatX \setminus \{v\})$.
In total, this gives us
$\cvalhatStX \geq \cvalhatStpXp + 1$.
Conversely, let $\hatXp$ be an extension of $X'$ at $t'$,
so $\scorehatStphatXp = \cvalhatStpXp $.
We distinguish two cases.
\begin{enumerate}
\item If $v \in X$, then
from
$v \notin N_G[\hatXp]_{t'}$ and
$v \in N_G[\hatXp \cup \{v\}]_t \cap \hatS$
we infer
$\scorehatSt(\hatXp \cup \{v\}) = \scorehatStphatXp + 1$.
Since $X = X' \cup \{v\}$ and $X' = \hatXp \cap \chi(t')$, it holds that
$X = (\hatXp \cup \{v\}) \cap \chi(t)$.
Hence the set $\hatXp \cup \{v\}$ is one of the candidates considered for an extension of $X$ in the definition of $\cvalhatSt$ and we obtain
$\cvalhatStX \leq \scorehatSt(\hatXp \cup \{v\})$.

\item Otherwise $v \notin X$.
In this case $X = X'$,
$v \notin \hatXp$ and
$X = \hatXp \cap \chi(t)$.
Hence the set $\hatXp$ is considered in the definition of $\cvalhatStX$ and we get
$\cvalhatStX \leq \scorehatSthatXp$.
Since $v$ is adjacent to an element of $X$,
we infer
$\scorehatSthatXp = \scorehatStphatXp + 1$.
\end{enumerate}
In both cases, we obtain
$\cvalhatStX \leq \cvalhatStpXp + 1$, so indeed
$\cvalhatStX = \cvalhatStpXp + 1$.

Next consider the case where $v \in N_G[X] \setminus \hatS$.
Clearly $v \notin X$.
Let $\hatX$ be an extension of $X$ at $t$,
so $\scorehatSthatX = \cvalhatStX$.
From
$v \notin N_G[\hatX]_{t'}$ and
$v \in N_G[\hatX]_t \setminus \hatS$
we now infer
$\scorehatSthatX = \scorehatStphatX - 1$.
Similar to before, by definition of $\cvalhatStpXp$ we obtain
$\cvalhatStpXp \leq \scorehatStphatX$.
In total, this gives us
$\cvalhatStX \geq \cvalhatStpXp - 1$.
Conversely, let $\hatXp$ be an extension of $X'$ at $t'$,
so $\scorehatStphatXp = \cvalhatStpXp $.
Since $v \notin \hatXp$ and
$X = \hatXp \cap \chi(t)$,
$\hatXp$ is considered in the definition of $\cvalhatStX$ and we get
$\cvalhatStX \leq \scorehatSthatXp$.
Since $v$ is adjacent to an element of $X$,
we infer
$\scorehatSthatXp = \scorehatStphatXp - 1$.
We obtain
$\cvalhatStX \leq \cvalhatStpXp - 1$, so indeed
$\cvalhatStX = \cvalhatStpXp - 1$.

Finally consider the remaining case where $v \notin N_G[X]$ and, in particular, $v \notin X$ holds as well as $X = X'$.
Using elementary set theory with
$\hatSt \setminus \{v\} = \hatStp$ and $\chi(t) = \chi(t') \cup \{v\}$ in mind,
we can prove that
$\{\hatX \subseteq \hatSt \mid \hatX \cap \chi(t) = X\}$
is equal to
$\{\hatX \subseteq \hatStp \mid \hatX \cap \chi(t') = X'\}$.
Hence a set $\hatX$ is considered in the definition of $\cvalhatStX$ iff
it is considered in the definition of $\cvalhatStpXp$.
For every $\hatX \subseteq \hatSt$ such that $\hatX \cap \chi(t) = X$,
observe that $v \notin N_G[\hatX]_t$, since $v$ is not adjacent to any element
of $X$ and if it were adjacent to some element of $\hatX \setminus X$,
then $\T$ would not be a valid tree decomposition.
This proves that every such $\hatX$ has the same score
at $t$ and $t'$.
Hence $\cvalhatStX = \cvalhatStpXp$.

\item[Forget node.]
Let $t$ be a forget node with child $t'$,
let $v$ be the unique element of $\chi(t') \setminus \chi(t)$ and
let $X \subseteq \hatS \cap \chi(t)$.
We prove that the following equation holds:
\[\cvalhatStX = \begin{cases}
\min\{\cvalhatStp(X),\; \cvalhatStp(X \cup \{v\})\} & \text{if } v \in \hatS\\
\cvalhatStp(X) & \text{otherwise}
\end{cases}\]
Clearly $\hatSt = \hatStp$ and all scores at forget nodes are identical to
those in the respective child node.
The case where $v \notin \hatS$ is trivial as then
$\hatS \cap \chi(t) = \hatS \cap \chi(t')$,
i.e., the domains of $\cvalhatSt$ and $\cvalhatStp$ are equal,
and the sets considered in the definitions of $\cvalhatStX$ and $\cvalhatStpX$
are the same.
Hence we consider the case where $v \in \hatS$.

Let $\hatX$ be an extension of $X$ at $t$,
so $\cvalhatStX = \scorehatSthatX = \scorehatStphatX$.
If $v \notin \hatX$, then
$\hatX \cap \chi(t') = X$, so we obtain
$\cvalhatStpX \leq \scorehatStphatX$ by definition of $\cvalhatStpX$.
On the other hand, if $v \in \hatX$, then
$\hatX \cap \chi(t') = X \cup \{v\}$, so we obtain
$\cvalhatStp(X \cup \{v\}) \leq \scorehatStphatX$.
As one of these two inequalities applies, we get
$\cvalhatStX \geq \min\{\cvalhatStpX,\; \cvalhatStp(X \cup \{v\})\}$.

Conversely, every extension $\hatXp$ of $X$ at $t'$
is considered in the definition of $\cvalhatStX$,
so $\cvalhatStX \leq \scorehatSthatXp = \scorehatStphatXp = \cvalhatStpX$.
Moreover, every extension $\hatXp$ of $X \cup \{v\}$ at $t'$
is also considered in the definition of $\cvalhatStX$,
so $\cvalhatStX \leq \scorehatSthatXp = \scorehatStphatXp = \cvalhatStp(X \cup \{v\})$.
If we combine these two inequalities, we get
$\cvalhatStX \leq \min\{\cvalhatStpX,\; \cvalhatStp(X \cup \{v\})\}$.
Hence
$\cvalhatStX = \min\{\cvalhatStpX,\; \cvalhatStp(X \cup \{v\})\}$.

\item[Join node.]
Let $t$ be a join node with children $t', t''$ such that
$\chi(t) = \chi(t') = \chi(t'')$,
and let $X \subseteq \hatS \cap \chi(t)$.
We prove that the following equation holds:
\[\cvalhatStX = \cvalhatStpX + \cvalhatStppX - \lscorehatStX\]
Let $\hatX$ be an extension of $X$ at $t$,
so $\scorehatSthatX = \cvalhatStX$.
The set $\hatXp = \hatX \cap \hatStp$ satisfies
$\hatXp \cap \chi(t') = X$, so
$\cvalhatStpX \leq \scorehatStphatXp$.
Symmetrically, for
$\hatXpp = \hatX \cap \hatStpp$ it holds that
$\cvalhatStppX \leq \scorehatStpphatXpp$.

There is no element of $V(G)_{t''} \setminus \chi(t)$
that is adjacent to an element of $\hatXp \setminus X$,
otherwise $\T$ would not be a valid tree decomposition.
Hence $N_G[\hatXp]_t = N_G[\hatXp]_{t'}$, and symmetrically
$N_G[\hatXpp]_t = N_G[\hatXpp]_{t''}$.
This entails
$\scorehatSthatXp = \scorehatStphatXp$ and
$\scorehatSthatXpp= \scorehatStpphatXpp$.

Since $N_G[\hatX]_t \cap \hatS$ is the union of $N_G[\hatXp]_t \cap \hatS$ and $N_G[\hatXpp]_t \cap \hatS$,
and these latter two sets have $N_G[X] \cap \chi(t) \cap \hatS$ as their intersection,
we can apply the inclusion-exclusion principle to obtain
$\size{N_G[\hatX]_t \cap \hatS} = \size{N_G[\hatXp]_t \cap \hatS} + \size{N_G[\hatXpp]_t \cap \hatS} - \size{N_G[X] \cap \chi(t) \cap \hatS}$.
In a similar way, we get
$\size{N_G[\hatX]_t \setminus \hatS} = \size{N_G[\hatXp]_t \setminus \hatS} + \size{N_G[\hatXpp]_t \setminus \hatS} - \size{(N_G[X] \cap \chi(t)) \setminus \hatS)}$.
We can establish
$\scorehatSthatX = \scorehatSthatXp + \scorehatSthatXpp - \lscorehatStX$
by putting these equations together.
The inequalities we have derived before now allow us to conclude
$\cvalhatStX \geq \cvalhatStpX + \cvalhatStppX - \lscorehatStX$.

Now let $\hatXp$ and $\hatXpp$ be extensions of $X$ at $t'$ and at $t''$, respectively.
We have that
$\cvalhatStpX = \scorehatStphatXp$ and
$\cvalhatStppX = \scorehatStpphatXpp$.
The set $\hatX = \hatXp \cup \hatXpp$
is clearly considered in the definition of $\cvalhatStX$,
so $\cvalhatStX \leq \scorehatSthatX$.
Following the same reasoning as before, we obtain
$\scorehatSthatX = \scorehatSthatXp + \scorehatSthatXpp - \lscorehatStX$.
This gives us
$\cvalhatStX \leq \cvalhatStpX + \cvalhatStppX - \lscorehatStX$.
Hence
$\cvalhatStX = \cvalhatStpX + \cvalhatStppX - \lscorehatStX$.
\end{description}

Using these recurrence relations, we can traverse the tree decomposition $\T$
in a bottom-up way and compute at each node $t$ of $\T$ the value $\cvalhatStX$
for each $X \subseteq \hatS \cap \chi(t)$.
Hence for each node of $\T$ we compute at most $2^w$ values, where $w$ is the
width of $\T$.
By choosing the right data structure for adjacency
tests~\cite[Exercise~7.16]{Cygan15},
each value can be computed in time $\O(w^3)$.
Since $\T$ has $\O(w \cdot \size{V(G)})$ many nodes and $\T$ can be computed in fixed-parameter
linear time~\cite{DBLP:journals/siamcomp/Bodlaender96}, (in fact in time $2^{\O(w^3)} \cdot 
\size{V(G)}$ as observed by \cite{stacs:BojanczykP17}),
we thus get an algorithm with fixed-parameter linear running time for checking
whether a given set $\hatS$ is secure.
\begin{theorem}
\label{thm:ssv-fpt-algorithm}
Given a graph $G$, a tree decomposition of $G$ of weight $w$ and a set $\hatS 
\subseteq V(G)$, we can decide  in time $\O(2^w \cdot w^4 \cdot 
\size{V(G)})$ whether $\hatS$ is secure in $G$.
\end{theorem}

Our algorithm can easily be adjusted to find a witness if $\hatS$ is not secure,
i.e., to print a subset of $\hatS$ that has more attackers than defenders.
After $\cvalhatSt$ has been computed for each $t$, this can be done via a final
top-down traversal by a standard technique in dynamic programming on tree
decompositions~\cite{niedermeier2006invitation}:
Alongside each value $\cvalhatStX$, we store the ``origin'' of this value and
recursively combine the origins of $\cval{\hatS}{r}(\emptyset)$, where $r$ is the
root of $\T$.

In our definition of the \SSV{} problem, we were only concerned with checking
whether a set is secure, but we did not mention the additional constructs that
we consider in this paper, like complementary vertex pairs or necessary or
forbidden vertices.
However, these additions pose no difficulty at all because we can just check
the respective conditions in linear time.

%
%

\subsection{A Polynomial Algorithm for Secure Set on Bounded Treewidth}

We now present an algorithm for finding secure sets, not just verifying
whether a given set is secure.
Our algorithm works by dynamic programming on a tree decomposition of the
input and extends the algorithm from
Section~\ref{sec:fpt-verification}.
For graphs of bounded treewidth, the algorithm presented in this section
runs in polynomial time.
However, in contrast to the algorithm in Section~\ref{sec:fpt-verification},
it is not an FPT algorithm since the degree of the polynomial depends on
the treewidth.
This is to be expected since the problem of finding secure sets of a certain
size is \Wone{}-hard when parameterized by treewidth, as we have shown in
Lemma~\ref{lem:ss-wone-hard}.
Our algorithm provides an upper bound for the complexity of this problem,
namely membership in the class \XP{}.

Let $G$ be a graph with a nice tree decomposition $\T$,
and let $t$ be a node of $\T$.
In Section~\ref{sec:fpt-verification}, we were given one particular secure set
candidate $\hatS$ that we wanted to check, so we only computed one value for each
$X \subseteq \hatS \cap \chi(t)$, namely the lowest score of any
$\hatX \subseteq \hatS_t$ whose intersection with $\chi(t)$ is $X$.
Here, in contrast, we cannot restrict ourselves to only one secure set
candidate, and multiple candidates may have the same intersection with
$\chi(t)$.
We therefore compute multiple objects for each subset of $\chi(t)$, since
two subsets of $V(G)_t$ that have the same intersection with $\chi(t)$
may have to be distinguished due to their subsets having different scores.

Let $S \subseteq \chi(t)$.
By $\possibleTablesS$ we denote the set of functions from $2^S$ to an integer.
Let $c \in \possibleTablesS$ and let
$k$ be an integer.
We say that a set $\hatS \subseteq V(G)_t$ is \emph{$(S,t,c,k)$-characterized} if
$\size{\hatS} = k$,
$\hatS \cap \chi(t) = S$ and,
for each
$X \subseteq S$, it holds that
$c(X) = \cval{\hatS}{t}(X)$, where $\cval{\hatS}{t}$ is the function defined
in Section~\ref{sec:fpt-verification}.
For each $S \subseteq \chi(t)$, we now define the set
\[\tablesSt = \{(c,k) \mid \text{there is a $(S,t,c,k)$-characterized set}\}.\]
When $r$ is the root of $\T$, there is a secure set of size $k$ in
$G$ if and only if there is an element $(c,k) \in \tables{\emptyset}{r}$ such
that
$c(\emptyset) \geq 0$.
To see this, first suppose there is a secure set $\hatS$ of size $k$
in $G$.
Then there is a function
$c: \{\emptyset\} \to \mathbb{Z}$ such that
$\hatS$ is $(\emptyset,r,c,k)$-characterized, so
$(c,k) \in \tables{\emptyset}{r}$ and
$c(\emptyset) = \cval{\hatS}{r}(\emptyset)$,
which means that $c(\emptyset)$ is the lowest score of any subset of $\hatS$.
Since $\hatS$ is secure in $G$, this number is nonnegative.
For the other direction,
let $(c,k) \in \tables{\emptyset}{r}$ such that
$c(\emptyset) \geq 0$.
Then there is a $(\emptyset,r,c,k)$-characterized set $\hatS$, obviously of
size $k$.
Since $c(\emptyset) \geq 0$,
the lowest score of any subset of $\hatS$ is nonnegative, which proves that
$\hatS$ is secure in $G$.

We now describe how to compute all such values in a bottom-up manner.

\begin{description}
\item[Leaf node.]
If $t$ is a leaf node, its bag is empty and obviously
$\tables{\emptyset}{t} = \{(c,0)\}$ holds, where $c$ maps $\emptyset$ to~$0$.

\item[Introduce node.]
Let $t$ be an introduce node with child $t'$ and
let $v$ be the unique element of $\chi(t) \setminus \chi(t')$.
For each $S \subseteq \chi(t)$ and
each function
$c \in \possibleTablesSsetminusv$,
we define a function
$c \plusS v \colon 2^S \to \mathbb{Z}$.
Its intended purpose is to obtain a version of $c$ that applies to $t$
instead of $t'$.
If $v \in S$, we need to increase scores where $v$ can serve as an additional
defender, and otherwise we need to decrease scores where $v$ can serve as an
additional attacker.
We now make this formal.
Let
$S \subseteq \chi(t)$,
$X \subseteq S$,
$X' = X \setminus \{v\}$ and
$c \in \possibleTablesSsetminusv$.
\[(c \plusS v)(X) = \begin{cases}
    c(X') + 1 &\text{if } v \in N_G[X] \cap S\\
    c(X') - 1 &\text{if } v \in N_G[X] \setminus S\\
    c(X') &\text{otherwise}
\end{cases}\]
For each $S \subseteq \chi(t)$
and each function $c \in \possibleTablesS$
there is a unique function
$c' \in \possibleTablesSsetminusv$
such that
$c = c' \plusS v$,
and we denote
$c'$ by $\origin(c)$.

The following statements can be proved by arguments similar to those in
Section~\ref{sec:fpt-verification}:
Let $\hatS \subseteq V(G)_{t'}$,
$S = \hatS \cap \chi(t')$ and
$(c,k) \in \tablesStp$
such that
$\hatS$ is $(S,t',c,k)$-characterized.
The set
$\hatS \cup \{v\}$ is $(S \cup \{v\},t,c \plusScupv v,k+1)$-characterized
and
$\hatS$ is $(S,t,c \plusS v,k)$-characterized.
Hence
$(c \plusScupv v,k+1) \in \tables{S \cup \{v\}}{t}$ and
$(c \plusS v,k) \in \tables{S}{t}$.
Conversely, let
$\hatS \subseteq V(G)_{t}$,
$S = \hatS \cap \chi(t)$ and
$(c,k) \in \tablesSt$
such that
$\hatS$ is $(S,t,c,k)$-characterized, and
let $c' = \origin(c)$
and $k' = k - \size{S \cap \{v\}}$.
The set $\hatS \setminus \{v\}$ is $(S \setminus \{v\},t',c',k')$-characterized.
Hence
$(c',k') \in \tables{S \setminus \{v\}}{t'}$.


From these observations, the following equation follows for every
$S \subseteq \chi(t)$:
\[\tablesSt = \{(c \plusS v, k+\size{S \cap \{v\}}) \mid (c,k) \in \tables{S \setminus \{v\}}{t'}\}\]

\item[Forget node.]
Let $t$ be a forget node with child $t'$ and
let $v$ be the unique element of $\chi(t') \setminus \chi(t)$.
For each $S \subseteq \chi(t)$ and
each function
$c \in \possibleTablesScupv$,
we define a function $c \minusS v$,
and for each $S \subseteq \chi(t)$ and
each function
$c \in \possibleTablesS$,
we define a function $c \minusNotS v$.
Each of these functions maps every subset of $S$ to an integer.
\[(c \minusS v)(X) = \min\{c(X),\; c(X \cup \{v\})\}\]
\[(c \minusNotS v)(X) = c(X)\]
Next we define functions $\originS$ and $\originNotS$ that map each
element of $\possibleTablesS$
to a set of elements of $\possibleTables{S \cup \{v\}}$ and $\possibleTablesS$, respectively:
\[\originSc = \{c' \in \possibleTablesScupv \mid c = c' \minusS v\}\]
\[\originNotSc = \{c' \in \possibleTablesS \mid c = c' \minusNotS v\}\]
The following statements can be proved by arguments similar to those in
Section~\ref{sec:fpt-verification}:
Let $\hatS \subseteq V(G)_{t'}$,
$S = \hatS \cap \chi(t')$ and
$(c,k) \in \tablesStp$ such that
$\hatS$ is $(S,t',c,k)$-characterized.
If $v \in \hatS$, then
$\hatS$ is $(S \setminus \{v\},t,c \minusS v,k)$-characterized and
$(c \minusS v,k) \in \tables{S \setminus \{v\}}{t}$;
otherwise
$\hatS$ is $(S,t,c \minusNotS v,k)$-characterized and
$(c \minusNotS v,k) \in \tables{S}{t}$.
Conversely, let $\hatS \subseteq V(G)_{t}$,
$S = \hatS \cap \chi(t)$ and
$(c,k) \in \tablesSt$ such that
$\hatS$ is $(S,t,c,k)$-characterized.
If $v \in \hatS$, then
there is some $c' \in \originSc$ such that
$\hatS$ is $(S \cup \{v\},t',c',k)$-characterized and
$(c',k) \in \tables{S \cup \{v\}}{t'}$;
otherwise
there is some $c' \in \originNotSc$ such that
$\hatS$ is $(S,t',c',k)$-characterized and
$(c',k) \in \tablesStp$.

From these observations, the following equation follows for every
$S \subseteq \chi(t)$:
\[\tablesSt =
\{(c \minusS v,k) \mid (c,k) \in \tables{S \cup \{v\}}{t'}\} \cup
\{(c \minusNotS v,k) \mid (c,k) \in \tables{S}{t'}\}\]

\item[Join node.]
Let $t$ be a join node with children $t', t''$ such that
$\chi(t) = \chi(t') = \chi(t'')$.
For each $S \subseteq \chi(t)$,
and each $c', c'' \in \possibleTablesS$,
we define a function $c' \joinSt c''$, which maps each subset of $S$ to an
integer.
\[(c' \joinSt c'')(X) = c'(X) + c''(X) - \lscoreStX\]
Next we define a function $\originJoinSt$ that maps each element of $\possibleTablesS$ to
a subset of $\possibleTablesS \times \possibleTablesS$:
\[\originJoinStc = \{(c',c'') \in \possibleTablesS \times \possibleTablesS \mid c = c' \joinSt c''\}\]
The following statements can be proved by arguments similar to those in
Section~\ref{sec:fpt-verification}:
Let
$\hatSp \subseteq V(G)_{t'}$,
$\hatSpp \subseteq V(G)_{t''}$,
$S = \hatSp \cap \hatSpp$,
$(c',k') \in \tablesStp$ and
$(c'',k'') \in \tablesStpp$ such that
$\hatSp$ is $(S,t',c',k')$-characterized and
$\hatSpp$ is $(S,t'',c'',k'')$-characterized,
and let
$c = c' \joinSt c''$ and
$k = k' + k'' - \size{S}$.
The set $\hatSp \cup \hatSpp$ is $(S,t,c,k)$-characterized
and $(c,k) \in \tablesSt$.
Conversely, let
$\hatS \subseteq V(G)_{t}$,
$S = \hatS \cap \chi(t)$ and
$(c,k) \in \tablesSt$ such that
$\hatS$ is $(S,t,c,k)$-characterized.
There is some $(c',c'') \in \originJoinStc$
as well as integers $k', k''$
such that
$k = k' + k'' - \size{S}$,
the set
$\hatS \cap V(G)_{t'}$ is $(S,t',c',k')$-characterized and
$\hatS \cap V(G)_{t''}$ is $(S,t'',c'',k'')$-characterized.
Hence $(c',k') \in \tablesStp$ and $(c'',k'') \in \tablesStpp$.

From these observations, the following equation follows for every
$S \subseteq \chi(t)$:
\[\tablesSt = \{(c' \joinSt c'', k' + k'' - \size{S}) \mid (c',k') \in \tablesStp,\; (c'',k'') \in \tablesStpp\}\]
\end{description}

We can now traverse the tree decomposition $\T$ in a bottom-up way and
at each node $t$ of $\T$ compute the set
$\tablesSt$ for each $S \subseteq \chi(t)$.
Let $n$ denote the number of vertices of $G$ and $w$ denote the width of $\T$.
Every element of $\tablesSt$ is a pair $(c,k)$,
where $c$ is a function that
maps each subset of $S$ to an integer between $-n$ and $n$,
there are at most $2^w$ subsets of $S$,
and $k$ is an integer between $0$ and $n$.
Hence there are at most
$(2n+1)^{2^w} \cdot (n+1)$ 
elements of $\tablesSt$.
Each individual element of $\tablesSt$ can be computed in time
$\O(2^w)$.
Finally, there are at most $2^w$ possible values for $S$ and $\O(wn)$ many nodes
in $\T$.
We thus get an algorithm that takes as input an integer $k$ together with a
graph $G$ whose treewidth we denote by $w$, and determines in time
$f(w) \cdot n^{g(w)}$ whether $G$ admits a secure set of size $k$,
where $f$ and $g$ are functions that only depend on $w$.

This algorithm for \ESS{} obviously also gives us an algorithm for \probSS{}
by checking all solution sizes from $1$ to $k$.
Finally, we can easily extend it to accommodate complementary vertex pairs as
well as necessary and forbidden vertices.
Hence we get the following \XP{} membership result:
\begin{theorem}
\label{thm:ss-xp-algorithm}
\probSS{},
\ESS{},
\SSF{},
\ESSF{},
\SSFN{},
\ESSFN{},
\SSFNC{} and
\ESSFNC{}
can be solved in polynomial time if the treewidth of the input is
bounded by a constant.
\end{theorem}

By keeping track of the origins of our computed values during our bottom-up
traversal of the tree decomposition, we can easily adapt the algorithm to find
solutions if they exist.

\section{Conclusion}
\label{sec:conclusion}

In this work, we have solved a complexity problem in graph theory that, to the
best of our knowledge, has remained open since the introduction of secure
sets~\cite{DBLP:journals/dam/BrighamDH07} in 2007.
We have shown that the problem of deciding whether, for a given graph $G$ and
integer $k$, $G$ possesses a non-empty secure set of size at most $k$ is
$\Sptwo$-complete.
We moreover obtained $\Sptwo$-completeness for seven further variants of this
problem.

In the second part of this paper, we analyzed the complexity of the \probSS{} 
problem parameterized by the treewidth of the input graph.
In particular, we showed that bounded treewidth does not make the problem 
fixed-parameter tractable unless $\FPT = \Wone$.
Nevertheless, we provided a polynomial-time algorithm for finding secure sets 
on graphs of bounded treewidth and thus showed membership in the class \XP.
As a positive result, we could show that the \CONP{}-complete problem of 
verifying whether a given set is secure can be solved in fixed-parameter linear 
time when parameterized by treewidth.

There are several interesting directions for future research.
One open question is which additional restrictions beside bounded treewidth 
need to be imposed on \probSS{} instances to achieve fixed-parameter 
tractability.
On the other hand, the \SSV{} problem may remain FPT for parameters that are 
less restrictive than treewidth.
We showed \Wone{}-hardness and \XP{}-membership of \probSS{}, so a tight bound 
is still lacking, albeit perhaps more of theoretical interest due to the fact 
that problems at a certain level of the weft hierarchy generally do not admit 
faster algorithms than problems at higher levels.
To classify a problem as FPT w.r.t.\ treewidth, a common approach is to express 
it in monadic second-order logic (MSO) and then invoke Courcelle's 
Theorem~\cite{DBLP:journals/iandc/Courcelle90}, which immediately proves that 
the problem is FPT.
We showed that \SSV{} is FPT, but it is not clear if it can be expressed in 
MSO.
If it cannot, then our FPT result could hint at possible extensions of MSO 
whose model-checking problem is still FPT.
Similarly, we believe that MSO can be extended in such a way that \probSS{} can 
be expressed and that a variant of Courcelle's Theorem for showing membership 
in \XP{} instead of \FPT{} holds.
Finally, some of our results seem to be transferable to (variants of) the 
\prob{Defensive Alliance} problem, so it would be interesting to investigate if 
some of our reductions and algorithms can help in the study of such related 
problems.


\bibliography{references}

\end{document}

%% file: secure-set-example-figure.tex
\begin{figure}
\centering
\begin{tikzpicture}
\node [draw,circle] (a) at (0.5,0) {$a$};
\node [draw,circle] (b) at (1.5,0) {$b$};
\node [draw,circle] (c) at (0,-1) {$c$};
\node (d) at (1,-1) {$d$};
\node (e) at (2,-1) {$e$};

\draw (a) -- (b);
\draw (c) -- (a) -- (d) -- (b) -- (e);
\draw (c) -- (d) -- (e);
\draw [bend right] (c) to (e);
\end{tikzpicture}
\caption{A graph with a minimum non-empty secure set indicated by circled vertices}
\label{fig:secure-set-example}
\end{figure}

%% file: secure-set-example-figure2.tex
\begin{figure}
\centering
\begin{tikzpicture}
\node [draw,necessary] (a) at (0,0) {$a$};
\node (b) at (1,0) {$b$};
\node at (1.5,0) {$\neq$};
\node (c) at (2,0) {$c$};
\node [necessary] (g) at (3,0) {};
\node (d) at (0,-1) {$d^\square$};
\node (e) at (1,-1) {$e$};
\node [forbidden] (f) at (2,-1) {};

\draw (a) -- (b);
\draw [bend left] (b) to (e);
\draw (e) -- (a);
\draw (d) -- (e) -- (f);
\draw (e) -- (c);
\draw (c) -- (g);

\path (b) -- (e) node [midway] {$\neq$};
\end{tikzpicture}
\caption{Illustration of forbidden, necessary and complementary vertices}
\label{fig:secure-set-example2}
\end{figure}

%% file: td-example-figure.tex
\begin{figure}[t]%
\centering
$G$:
\begin{tikzpicture}[scale=0.75, baseline=(d.base)]
\node (a) at (0,1) {$a$};
\node (b) at (-1,0) {$b$};
\node (c) at (0,-1) {$c$};
\node (d) at (1,0) {$d$};
\draw (a) -- (b) -- (c) -- (d) -- (a) -- (c);
\end{tikzpicture}%
\hspace{2em}
$\T$:
\begin{tikzpicture}[level/.style={sibling distance=8mm,level distance=12mm},
level 1/.style={level distance=8mm},
grow'=right, font=\footnotesize, anchor=west, growth parent anchor=west,
baseline=(root.base)]
\node (root) {$\emptyset$}
	child {node [label=above:$t^\T_a$] {$\{a\}$}
		child {node [label=above:$t^\T_c$] {$\{a,c\}$}
			child {node {$\{a,c\}$}
				child {node [label=above:$t^\T_b$] {$\{a,b,c\}$}
					child {node {$\{a,b\}$}
						child {node {$\{a\}$}
							child {node {$\emptyset$}}
						}
					}
				}
			}
			child {node {$\{a,c\}$}
				child {node [label=below:$t^\T_d$] {$\{a,c,d\}$}
					child {node {$\{c,d\}$}
						child {node {$\{d\}$}
							child {node {$\emptyset$}}
						}
					}
				}
			}
		}
	}
	;
\end{tikzpicture}%
\caption{A graph $G$ and a nice tree decomposition $\T$ of $G$ rooted at the leftmost node}
\label{fig:td-example}
\end{figure}

%% file: qsat-reduction-figure.tex
\begin{figure}
\centering
\begin{tikzpicture}[yscale=0.4,xscale=1.25,looseness=0.8]
\node (nt1s) at (0,-2) {$\overline{t_1}^\square$};
\node (nt2s) at (0,-7) {$\overline{t_2}^\square$};
\node (nt3s) at (0,-12) {$\overline{t_3}^\square$};

\node (x1) at (1,-1) {$x_1$};
\node at (1,-2) {$\neq$};
\node (nx1) at (1,-3) {$\overline{x_1}$};
\node (x2) at (1,-6) {$x_2$};
\node at (1,-7) {$\neq$};
\node (nx2) at (1,-8) {$\overline{x_2}$};
\node (x3) at (1,-11) {$x_3$};
\node at (1,-12) {$\neq$};
\node (nx3) at (1,-13) {$\overline{x_3}$};

\node (d1s) at (1,-16) {$d_1^\square$};
\node (d2s) at (2,-16) {$d_2^\square$};

\node (nt1t) [necessary,yshift=1mm] at (2,-1) {};
\node (nt1) at (2,-2) {$\overline{t_1}$};
\node (nt2t) [necessary,yshift=1mm] at (2,-6) {};
\node (nt2) at (2,-7) {$\overline{t_2}$};
\node (nt3t) [necessary,yshift=1mm] at (2,-11) {};
\node (nt3) at (2,-12) {$\overline{t_3}$};

\node (nts) at (3,-16) {$\overline{t}^\square$};

\node at (2.5,-2) {$\neq$};
\node (t1) at (3,-2) {$t_1'$};
\node (t1s) [forbidden,yshift=1mm] at (3,-1) {};
\node at (2.5,-7) {$\neq$};
\node (t2) at (3,-7) {$t_2'$};
\node (t2s) [forbidden,yshift=1mm] at (3,-6) {};
\node at (2.5,-12) {$\neq$};
\node (t3) at (3,-12) {$t_3'$};
\node (t3s) [forbidden,yshift=1mm] at (3,-11) {};

\node at (2,-3) {$\neq$};
\node (t1p) at (2,-4) {$t_1$};
\node at (2,-8) {$\neq$};
\node (t2p) at (2,-9) {$t_2$};
\node at (2,-13) {$\neq$};
\node (t3p) at (2,-14) {$t_3$};

\node at (3,-3) {$\neq$};
\node (nt1p) at (3,-4) {$\overline{t_1'}$};
\node (nt1ps) [forbidden] at (2.5,-4) {};
\node at (3,-8) {$\neq$};
\node (nt2p) at (3,-9) {$\overline{t_2'}$};
\node (nt2ps) [forbidden] at (2.5,-9) {};
\node at (3,-13) {$\neq$};
\node (nt3p) at (3,-14) {$\overline{t_3'}$};
\node (nt3ps) [forbidden] at (2.5,-14) {};

\node (y1) [necessary] at (6,-2) {$y_1$};
\node (ny1) [necessary] at (6,-6) {$\overline{y_1}$};
\node (y2) [necessary] at (6,-10) {$y_2$};
\node (ny2) [necessary] at (6,-14) {$\overline{y_2}$};

\node (y1s1) [forbidden] at (6,-3.7) {};
\node (y1s2) [forbidden] at (6.2,-3.7) {};
\node (y1s3) [forbidden] at (6.4,-3.7) {};
\node (y1s4) [forbidden] at (6.6,-3.7) {};

\node (y1t1) [necessary] at (6.2,-1) {};
\node (y1t2) [necessary] at (6.4,-1) {};
\node (y1t3) [necessary] at (6.6,-1) {};

\node (ny1t1) [necessary] at (6,-7) {};
\node (ny1t2) [necessary] at (6.2,-7) {};
\node (ny1t3) [necessary] at (6.4,-7) {};

\node (y2s1) [forbidden] at (6,-11.7) {};
\node (y2s2) [forbidden] at (6.2,-11.7) {};
\node (y2s3) [forbidden] at (6.4,-11.7) {};
\node (y2s4) [forbidden] at (6.6,-11.7) {};

\node (y2t1) [necessary] at (6.2,-9) {};
\node (y2t2) [necessary] at (6.4,-9) {};
\node (y2t3) [necessary] at (6.6,-9) {};

\node (ny2t1) [necessary] at (6,-15) {};
\node (ny2t2) [necessary] at (6.2,-15) {};
\node (ny2t3) [necessary] at (6.4,-15) {};

\node (yt1) [necessary] at (8,-8) {};
\node (yt2) [necessary] at (8.2,-8) {};

\draw (nt1s) -- (x1) -- (nt1);
\draw (nt1s) -- (nx2) -- (nt1);
\draw (nt2s) -- (nx3) -- (nt2);
\draw (nt3s) -- (nx3) -- (nt3);
\draw [nearly transparent,dashed] (d1s) -- (nt3);
\draw [nearly transparent,dashed] (d1s) -- (nt2);

\draw (nt1) -- (nt1t);
\draw (nt2) -- (nt2t);
\draw (nt3) -- (nt3t);

\draw [nearly transparent,dashed] (nts) -- (nt1);
\draw [nearly transparent,dashed] (nts) -- (nt2);
\draw [nearly transparent,dashed] (nts) -- (nt3);

\draw (t1s) -- (t1);
\draw (t2s) -- (t2);
\draw (t3s) -- (t3);

\draw (nt1p) -- (nt1ps);
\draw (nt2p) -- (nt2ps);
\draw (nt3p) -- (nt3ps);

\draw (y1) -- (y1s1) -- (ny1);
\draw (y1) -- (y1s2) -- (ny1);
\draw (y1) -- (y1s3) -- (ny1);
\draw (y1) -- (y1s4) -- (ny1);

\draw (y1) -- (y1t1);
\draw (y1) -- (y1t2);
\draw (y1) -- (y1t3);

\draw (ny1) -- (ny1t1);
\draw (ny1) -- (ny1t2);
\draw (ny1) -- (ny1t3);

\draw (y2) -- (y2s1) -- (ny2);
\draw (y2) -- (y2s2) -- (ny2);
\draw (y2) -- (y2s3) -- (ny2);
\draw (y2) -- (y2s4) -- (ny2);

\draw (y2) -- (y2t1);
\draw (y2) -- (y2t2);
\draw (y2) -- (y2t3);

\draw (ny2) -- (ny2t1);
\draw (ny2) -- (ny2t2);
\draw (ny2) -- (ny2t3);

\foreach \t in {(t1),(t2),(t3)}
  \foreach \y in {(y1),(ny1),(y2),(ny2)}
    \draw [nearly transparent,dashed] \t -- \y;

\draw (nt1p) -- (ny1);
\draw (nt2p) -- (y1);
\draw (nt2p) -- (ny2);
\draw (nt3p) -- (y1);
\draw (nt3p) -- (y2);

\draw [bend left] (y1) to (yt1);
\draw [bend left] (y1) to (yt2);
\draw [bend left] (ny1) to (yt1);
\draw [bend left] (ny1) to (yt2);
\draw [bend right] (y2) to (yt1);
\draw [bend right] (y2) to (yt2);
\draw [bend right] (ny2) to (yt1);
\draw [bend right] (ny2) to (yt2);

\node [anchor=west,font=\footnotesize] at (y1t3.east) {($n_t$)};
\node [anchor=west,font=\footnotesize,yshift=-3mm,xshift=-2mm] at (y1s4.east) {($n_t+1$)};
\node [anchor=west,font=\footnotesize] at (yt2.east) {($n_t-1$)};
\end{tikzpicture}
\caption{Graph corresponding to the \prob{Qsat$_2$} formula $\exists x_1 \exists x_2 \exists x_3\; \forall y_1 \forall y_2\; \big((\neg x_1 \land x_2 \land y_1) \lor (x_3 \land \neg y_1 \land y_2) \lor (x_3 \land \neg y_1 \land \neg y_2)\big)$.
To avoid clutter, we omit labels for the vertices from $\YT$, $\YPT$, $\YS$, $\NTT$, $\TPS$ and $\NTPS$, and we draw some edges in a dashed style.}
\label{fig:qsat-reduction-example}
\end{figure}

%% file: complementary-reduction-figure.tex
\begin{figure}
\centering
\begin{tikzpicture}[xscale=1.09, yscale=0.9] 
\node (a) at (1.25,0) {$a$};
\node (ap) at (4.5,0) {$a^{ab}$};

\node (a1) at (0,-1) {$a_1^{ab}$};
\node (a2) at (1,-1) {$a_2^{ab}$};
\node at (1.75,-1) {$\cdots$};
\node (an1) at (2.5,-1) {$a_{n+1}^{ab}$};
\node (an2) at (3.5,-1) {$a_{n+2}^{ab}$};
\node (an3) at (4.5,-1) {$a_{n+3}^{ab}$};
\node (an4) at (5.5,-1) {$a_{n+4}^{ab}$};

\node (a1s) at (0,-2) {$a_1^{ab\square}$};
\node (a2s) at (1,-2) {$a_2^{ab\square}$};
\node at (1.75,-2) {$\cdots$};
\node (an1s) at (2.5,-2) {$a_{n+1}^{ab\square}$};
\node (an2s) at (3.5,-2) {$a_{n+2}^{ab\square}$};
\node (an3s) at (4.5,-2) {$a_{n+3}^{ab\square}$};
\node (an4s) at (5.5,-2) {$a_{n+4}^{ab\square}$};

\node (triangle) at (6,0) {$\triangle^{ab}$};

\node (bp) at (7.5,0) {$b^{ab}$};
\node (b) at (10.75,0) {$b$};

\node (bn4) at (6.5,-1) {$b_{n+4}^{ab}$};
\node (bn3) at (7.5,-1) {$b_{n+3}^{ab}$};
\node (bn2) at (8.5,-1) {$b_{n+2}^{ab}$};
\node (bn1) at (9.5,-1) {$b_{n+1}^{ab}$};
\node (bn) at (10.5,-1) {$b_n^{ab}$};
\node at (11.25,-1) {$\cdots$};
\node (b1) at (12,-1) {$b_1^{ab}$};

\node (bn4s) at (6.5,-2) {$b_{n+4}^{ab\square}$};
\node (bn3s) at (7.5,-2) {$b_{n+3}^{ab\square}$};
\node (bn2s) at (8.5,-2) {$b_{n+2}^{ab\square}$};
\node (bn1s) at (9.5,-2) {$b_{n+1}^{ab\square}$};
\node (bns) at (10.5,-2) {$b_n^{ab\square}$};
\node at (11.25,-2) {$\cdots$};
\node (b1s) at (12,-2) {$b_1^{ab\square}$};


\draw (a) -- (a1);
\draw (a) -- (a2);
\draw (a) -- (an1);
\draw (ap) -- (an2);
\draw (ap) -- (an3);
\draw (ap) -- (an4);
\draw (ap) -- (triangle) -- (bp);
\draw (b) -- (b1);
\draw (b) -- (bn1);
\draw (b) -- (bn);
\draw (bp) -- (bn2);
\draw (bp) -- (bn3);
\draw (bp) -- (bn4);

\draw (a1) -- (a2);
\draw (an1) -- (an2) -- (an3) -- (an4);
\draw (bn) -- (bn1) -- (bn2) -- (bn3) -- (bn4);

\draw (an1) -- (an1s);
\draw (an2) -- (an2s);
\draw (an3) -- (an3s);
\draw (an4) -- (an4s);
\draw (b1) -- (b1s);
\draw (bn) -- (bns);
\draw (bn1) -- (bn1s);
\draw (bn2) -- (bn2s);
\draw (bn3) -- (bn3s);
\draw (bn4) -- (bn4s);

\draw (a1) -- (a1s) -- (a2) -- (a2s);
\draw (a1) -- (a2s);
\draw (an1) -- (an2s) -- (an3) -- (an4s);
\draw (an1s) -- (an2) -- (an3s) -- (an4);
\draw (bns) -- (bn1) -- (bn2s) -- (bn3) -- (bn4s);
\draw (bn) -- (bn1s) -- (bn2) -- (bn3s) -- (bn4);
\end{tikzpicture}
\caption{Gadget for a pair of complementary vertices $(a,b)$ in the reduction from \SSFNC{} to \SSFN{}. The vertices $a$ and $b$ may have additional neighbors from the original graph.}
\label{fig:complementary-reduction-example}
\end{figure}

%% file: necessary-reduction-example-figure.tex
\begin{figure}
\centering
\begin{tikzpicture}[scale=0.7,inner sep=2pt]
\node (a) at (2,4) {$a$};

\node (a1) at (1,5) [anonymous] {};
\node (a2) at (1,4) [anonymous] {};
\node (a3) at (1,3) [anonymous] {};

\node (a1s) at (0,5) [forbidden] {};
\node (a2s) at (0,4) [forbidden] {};
\node (a3s) at (0,3) [forbidden] {};

\node (wa) at (1,2) {$w_a$};
\node (wap) at (1,1) {$w_a'$};
\node (w) at (3,1) {$w$};
\node (was) at (0,2) [forbidden] {};
\node (waps) at (0,1) [forbidden] {};

\node (w1) at (4,2) [anonymous] {};
\node (w2) at (4,1) [anonymous] {};
\node (w3) at (4,0) [anonymous] {};
\node (w1s) at (5,2) [forbidden] {};
\node (w2s) at (5,1) [forbidden] {};
\node (w3s) at (5,0) [forbidden] {};

\node (b) at (3,4) {$b$};

\node (b1) at (4,5) [anonymous] {};
\node (b2) at (4,4) [anonymous] {};
\node (b3) at (4,3) [anonymous] {};

\node (b1s) at (5,5) [forbidden] {};
\node (b2s) at (5,4) [forbidden] {};
\node (b3s) at (5,3) [forbidden] {};

\draw (a) -- (b);

\draw (a) -- (a1);
\draw (a) -- (a2);
\draw (a) -- (a3);

\draw (w) -- (wa);
\draw (w) -- (wap);

\draw (w) -- (w1);
\draw (w) -- (w2);
\draw (w) -- (w3);

\draw (a1) -- (a1s);
\draw (a2) -- (a2s);
\draw (a3) -- (a3s);
\draw (wa) -- (was);

\draw (a1) -- (a2) -- (a3) -- (wa) -- (wap);
\draw (a1) -- (a2s) -- (a3) -- (was);
\draw (a1s) -- (a2) -- (a3s) -- (wa);
\draw (waps) -- (wa);

\draw (b) -- (b1);
\draw (b) -- (b2);
\draw (b) -- (b3);

\draw (b1) -- (b1s);
\draw (b2) -- (b2s);
\draw (b3) -- (b3s);
\draw (w1) -- (w1s);
\draw (w2) -- (w2s);
\draw (w3) -- (w3s);

\draw (b1) -- (b2) -- (b3) -- (w1) -- (w2) -- (w3);
\draw (b1) -- (b2s) -- (b3) -- (w1s) -- (w2) -- (w3s);
\draw (b1s) -- (b2) -- (b3s) -- (w1) -- (w2s) -- (w3);
\end{tikzpicture}
\caption{Result of the transformation $\tFN{}$ applied to an example graph with two adjacent vertices $a$ and $b$, where $b$ is necessary. Every solution in the depicted graph contains $b$, $w$ and $w_a'$.}
\label{fig:necessary-reduction-example-figure}
\end{figure}

%% file: main.bbl
\begin{thebibliography}{10}

\bibitem{jlc:AbseherBCDW15}
Michael Abseher, Bernhard Bliem, G\"unther Charwat, Frederico Dusberger, and
  Stefan Woltran.
\newblock Computing secure sets in graphs using answer set programming.
\newblock In {\em J. Logic Comput.}, 2015.
\newblock Accepted for publication.

\bibitem{Arnborg:1987:CFE:37170.37183}
Stefan Arnborg, Derek~G. Corneil, and Andrzej Proskurowski.
\newblock Complexity of finding embeddings in a k-tree.
\newblock {\em SIAM J. Algebraic Discrete Methods}, 8(2):277--284, 1987.

\bibitem{DBLP:journals/dam/AsahiroMO11}
Yuichi Asahiro, Eiji Miyano, and Hirotaka Ono.
\newblock Graph classes and the complexity of the graph orientation minimizing
  the maximum weighted outdegree.
\newblock {\em Discrete Applied Mathematics}, 159(7):498--508, 2011.

\bibitem{DBLP:conf/wg/BliemW15}
Bernhard Bliem and Stefan Woltran.
\newblock Complexity of secure sets.
\newblock In {\em {WG}}, volume 9224 of {\em Lecture Notes in Computer
  Science}, pages 64--77. Springer, 2015.

\bibitem{DBLP:journals/actaC/Bodlaender93}
Hans~L. Bodlaender.
\newblock A tourist guide through treewidth.
\newblock {\em Acta Cybernet.}, 11(1-2):1--22, 1993.

\bibitem{DBLP:journals/siamcomp/Bodlaender96}
Hans~L. Bodlaender.
\newblock A linear-time algorithm for finding tree-decompositions of small
  treewidth.
\newblock {\em SIAM J. Comput.}, 25(6):1305--1317, 1996.

\bibitem{DBLP:conf/sofsem/Bodlaender05}
Hans~L. Bodlaender.
\newblock Discovering treewidth.
\newblock In {\em Proc.\ SOFSEM}, volume 3381 of {\em LNCS}, pages 1--16.
  Springer, 2005.

\bibitem{DBLP:journals/iandc/BodlaenderK10}
Hans~L. Bodlaender and Arie M. C.~A. Koster.
\newblock Treewidth computations {I}. {U}pper bounds.
\newblock {\em Inf. Comput.}, 208(3):259--275, 2010.

\bibitem{stacs:BojanczykP17}
Miko{\l}aj Boja{\'{n}}czyk and Michal Pilipczuk.
\newblock Optimizing tree decompositions in {MSO}.
\newblock In {\em Proc.\ {STACS}}, volume~66 of {\em LIPIcs}, pages
  15:1--15:13. Schloss Dagstuhl - Leibniz-Zentrum fuer Informatik, 2017.

\bibitem{DBLP:journals/cacm/BrewkaET11}
Gerhard Brewka, Thomas Eiter, and Miros{\l}aw Truszczy{\'n}ski.
\newblock Answer set programming at a glance.
\newblock {\em Commun. ACM}, 54(12):92--103, 2011.

\bibitem{DBLP:journals/dam/BrighamDH07}
Robert~C. Brigham, Ronald~D. Dutton, and Stephen~T. Hedetniemi.
\newblock Security in graphs.
\newblock {\em Discrete Appl. Math.}, 155(13):1708--1714, 2007.

\bibitem{DBLP:journals/iandc/Courcelle90}
Bruno Courcelle.
\newblock The monadic second-order logic of graphs. {I}. {R}ecognizable sets of
  finite graphs.
\newblock {\em Inf. Comput.}, 85(1):12--75, 1990.

\bibitem{Cygan15}
Marek Cygan, Fedor~V. Fomin, Lukasz Kowalik, Daniel Lokshtanov, D{\'{a}}niel
  Marx, Marcin Pilipczuk, Michal Pilipczuk, and Saket Saurabh.
\newblock {\em Parameterized Algorithms}.
\newblock Springer, 2015.

\bibitem{DBLP:conf/micai/DermakuGGMMS08}
Artan Dermaku, Tobias Ganzow, Georg Gottlob, Benjamin~J. McMahan, Nysret
  Musliu, and Marko Samer.
\newblock Heuristic methods for hypertree decomposition.
\newblock In {\em Proc.\ MICAI}, volume 5317 of {\em LNCS}, pages 1--11.
  Springer, 2008.

\bibitem{DBLP:conf/iwpec/DomLSV08}
Michael Dom, Daniel Lokshtanov, Saket Saurabh, and Yngve Villanger.
\newblock Capacitated domination and covering: {A} parameterized perspective.
\newblock In {\em Proc.\ {IWPEC}}, volume 5018 of {\em LNCS}, pages 78--90.
  Springer, 2008.

\bibitem{downey1999parameterized}
Rodney~G. Downey and Michael~R. Fellows.
\newblock {\em Parameterized Complexity}.
\newblock Monographs in Computer Science. Springer, 1999.

\bibitem{DuttonE08}
Rosa~I. Enciso and Ronald~D. Dutton.
\newblock Parameterized complexity of secure sets.
\newblock In {\em Congr. Numer.}, volume 189, pages 161--168, 2008.

\bibitem{DBLP:journals/computer/FlakeLGC02}
Gary~William Flake, Steve Lawrence, C.~Lee Giles, and Frans Coetzee.
\newblock Self-organization and identification of web communities.
\newblock {\em {IEEE} Computer}, 35(3):66--71, 2002.

\bibitem{flum2006parameterized}
J{\"o}rg Flum and Martin Grohe.
\newblock {\em Parameterized Complexity Theory}.
\newblock Texts in Theoretical Computer Science. Springer, 2006.

\bibitem{DBLP:journals/combinatorics/HaynesHH03}
Teresa~W. Haynes, Stephen~T. Hedetniemi, and Michael~A. Henning.
\newblock Global defensive alliances in graphs.
\newblock {\em Electr. J. Comb.}, 10, 2003.

\bibitem{Ho2011}
Yiu~Yu Ho.
\newblock {\em Global Secure Sets of Trees and Grid-like Graphs}.
\newblock PhD thesis, University of Central Florida, Orlando, Florida, USA,
  2011.

\bibitem{ho2009rooted}
Yiu~Yu Ho and Ronald~D. Dutton.
\newblock Rooted secure sets of trees.
\newblock {\em AKCE Int. J. Graphs Comb.}, 6(3):373--392, 2009.

\bibitem{Kloks1994treewidth}
Ton Kloks.
\newblock {\em Treewidth: Computations and Approximations}, volume 842 of {\em
  LNCS}.
\newblock Springer, 1994.

\bibitem{dam:KornaiT92}
Andr{\'{a}}s Kornai and Zsolt Tuza.
\newblock Narrowness, pathwidth, and their application in natural language
  processing.
\newblock {\em Discrete Applied Mathematics}, 36(1):87--92, 1992.

\bibitem{DBLP:journals/tcs/Marx11}
D{\'{a}}niel Marx.
\newblock Complexity of clique coloring and related problems.
\newblock {\em Theor. Comput. Sci.}, 412(29):3487--3500, 2011.

\bibitem{niedermeier2006invitation}
Rolf Niedermeier.
\newblock {\em Invitation to Fixed-Parameter Algorithms}.
\newblock Oxford Lecture Series in Mathematics and its Applications. OUP, 2006.

\bibitem{DBLP:journals/jct/RobertsonS84}
Neil Robertson and Paul~D. Seymour.
\newblock Graph minors. {III}. {P}lanar tree-width.
\newblock {\em J. Comb. Theory, Ser. B}, 36(1):49--64, 1984.

\bibitem{DBLP:journals/amai/Szeider05}
Stefan Szeider.
\newblock Generalizations of matched {CNF} formulas.
\newblock {\em Ann. Math. Artif. Intell.}, 43(1):223--238, 2005.

\bibitem{DBLP:journals/corr/abs-1107-1177}
Stefan Szeider.
\newblock Not so easy problems for tree decomposable graphs.
\newblock {\em CoRR}, abs/1107.1177, 2011.

\bibitem{iandc:Thorup98}
Mikkel Thorup.
\newblock All structured programs have small tree-width and good register
  allocation.
\newblock {\em Inform. and Comput.}, 142(2):159--181, 1998.

\bibitem{yero2013defensive}
Ismael~Gonz{\'a}lez Yero and Juan~A Rodr{\'\i}guez-Vel{\'a}zquez.
\newblock Defensive alliances in graphs: a survey.
\newblock {\em CoRR}, abs/1308.2096, 2013.

\end{thebibliography}
